\documentclass[useAMS,usenatbib]{mn2e}

\usepackage{natbib}
\usepackage{multirow}
\usepackage{graphicx}
\usepackage{longtable}
\usepackage{amsmath}
\usepackage{subfigure}

\title[White Dwarfs in the SuperCOSMOS Sky Survey]{White Dwarfs in the SuperCOSMOS Sky Survey: the Thin Disk, Thick Disk and Spheroid Luminosity Functions}
\author[N. Rowell and N. C. Hambly]{N. Rowell$^{1}$\thanks{E-mail:
nickrowell@computing.dundee.ac.uk} and N. C.
Hambly$^{2}$\\
$^{1}$Space Technology Centre, School of Computing,\\
~University of Dundee, Dundee, DD1 4HN\\
$^{2}$Scottish Universities Physics Alliance (SUPA), Institute for Astronomy,\\
~School of Physics and Astronomy, University of Edinburgh,\\
~Royal Observatory, Blackford Hill, Edinburgh EH9 3HJ}
\begin{document}

\date{Accepted xx.xx.xxxx. Received xx.xx.xxxx; in original form xx.xx.xxxx}

\pagerange{\pageref{firstpage}--\pageref{lastpage}} \pubyear{2002}

\maketitle

\begin{abstract}
We present a magnitude and proper motion limited catalogue of $\sim10,000$ white dwarf candidates, 
obtained from the SuperCOSMOS Sky Survey by means of reduced proper motion selection.
This catalogue extends to magnitudes $R\sim19.75$ and proper motions as low as $\mu \sim 0\farcs05$ yr$^{-1}$, and covers 
nearly three quarters of the sky.
Photometric parallaxes provide distance estimates accurate to $\sim 50\%$.
This catalogue is used to measure the luminosity functions for disk and spheroid white dwarfs, using
strict velocity cuts to isolate subsamples belonging to each population. Disk luminosity functions measured in 
this manner are really a conglomerate of thin and thick disk objects, due to the significant velocity
overlap between these populations. We introduce a new statistical approach to the stellar luminosity
function for nearby objects that succesfully untangles the contributions from the different kinematic populations,
without the need for stringent velocity cuts. This improves the statistical power by allowing all stars to
contribute to the luminosity function, even at tangential velocities where the populations are indistinguishable.
This method is particularly suited to white dwarfs, for which
population discrimination by chemical tagging is not possible. We use this technique to obtain the first
measurement of the thick disk white dwarf luminosity function, while also improving constraint on both
the thin disk and spheroid luminosity functions. We find that the thin disk, thick disk and spheroid populations
contribute to the local white dwarf density in roughly 79\%/16\%/5\% proportions.
\end{abstract}

\section{Introduction}
\label{one}
Nearly all stars in the Galaxy will end their lives as faint white dwarfs (WDs), once nuclear burning has ceased and
the outer photospheric layers have been ejected. The further evolution of these objects is a gradual 
cooling process, where the remaining thermal energy of the exposed core is slowly radiated away through the thin outer layers.
WDs are highly numerous in the Solar neighbourhood. They are the second most common type of star behind low mass main sequence stars,
but due to their extreme faintness are found in much smaller numbers in Galactic star surveys. As a result, the evolution of the
coolest WDs is rather poorly constrained observationally
%
\citep{hansen2003,gates2004,vidrih2007,rowell2008,kilic2010UCWDs},
and though much work has been done on the theoretical front in recent years
%
\citep{fontaine2001,BLR2001}
it is still unclear how to interpret the small amount of observational evidence available
%
%
\citep[e.g.~][]{bergeron2002,bedin2008a}.
Studies of the luminosity function for white dwarfs (WDLF) typically focus on the faint end, where the finite age of the Galaxy
predicts a sudden downturn in the local density of WDs when the cooling times converge on the age of the population as a whole.
The first evidence for this was published by \citet{liebert1988}, and was used by \citet{winget1987}
to obtain an estimate of the age of the universe essentially independent of main sequence stellar ages.
Although many more studies have been carried out in the intervening decades, focussing on both the Galactic disk
\citep{oswalt1996,legget1998,knox1999}
and clusters
\citep{hansen2002,bedin2010}
 it is the advent of digitized wide angle sky surveys in the last decade
that has revolutionised the study of these objects in the field, 
with large footprint areas greatly increasing the survey volume for faint objects.

The Sloan Digital Sky Survey (SDSS)
%
%
has provided the most interesting results in recent years. Several independent surveys have increased the catalogue of known WDs by more than an
order of magnitude, providing greater statistical power at both the hot end of the LF, where the WDs are intrinsically rare, and the cool end where
they exist in large number but are extremely faint.
For example, \cite{krzesinski2009} use colour selection to identify a few hundred hot WDs from the SDSS spectral catalogue. Their luminosity
function extends to $M_{bol}=7.0$, where the sample starts to become incomplete due to the selection probability for spectroscopic follow up 
dropping off at lower temperatures.
\cite{eisenstein2006} identified more than 9,000 spectroscopically confirmed WDs in SDSS DR4, and their catalogue was used by \cite{degennaro2008}
to probe the WDLF for different WD mass ranges. As cool WDs are generally selected for SDSS spectroscopic follow-up by algorithms 
designed to target QSOs, the completeness of such a catalogue is a strong function of colour.
\citet{kilic2006} and \citet{harris2006} (hereafter H06) used reduced proper motion to identify $\sim6000$ WDs, with proper motions determined by pairing sources with
the USNO-B catalogue. The reduced proper motion approach offers a superior survey method for a couple of reasons. 
First, it is insensitive to colour, and cool WDs can be identified
among the highly numerous disk stars from which they are indistiguishable in colour space. Second, it provides a statistically rigorous 
selection probability, avoiding the need to calculate detailed completeness functions to prevent survey bias.
The WDLF of H06 is currently the deepest and best constrained around the faint turnover in the disk LF, though
the pairing criterion limits the depth of their catalogue to the magnitude limits of the USNO-B survey.

There has been considerable interest in Galactic spheroid WDs in the last decade, though very little is known about this elusive
population of stars. Broad predictions can be made on the basis of Galactic formation and stellar evolution models,
%
\citep[e.g.~][]{hansen1999,reid2005}
but the contribution made by these objects to the local density and Galactic mass budget is very poorly constrained.
This is due to a combination of several factors. Firstly, these objects are both faint and intrinsically very rare,
and in any given survey will be found in much smaller numbers than both disk WDs and other types of star.
They are far too faint to be seen at large Galactic plane distances in regions of pure spheroid stars, and any objects
must be observed as they pass through the nearby Galactic disk. They also cannot be identified by low metallicities, because
for all WDs the surface gravity is high enough to pull metals below the photosphere. The only reliable way to 
identify them is by their large space velocities, as the high velocity tail of the distribution is cleanly
separated from both the thin and thick disk \citep{vidrih2007,kilic2010halo}.
This technique was used by H06 to measure the first
luminosity function for spheroid WDs of any significant size.

The thin and thick disk WD populations overlap too much in velocity for single objects to be reliably attributed to either
population. For this reason, there has never been a direct measurement of the thick disk WDLF, though all previous
measurements of the WDLF for thin disk stars will have contained a significant number of these objects. In this
paper, we show that the thin and thick disk WDLFs can indeed be measured separately from a mixed catalogue, 
by avoiding the need to classify each survey object as one type or the other. This technique has more statistical
power than the traditional approach, because all objects are allowed to contribute to the luminosity function, 
even at tangential velocities where the populations are indistinguishable.

This paper is organised as follows: in Section \ref{two} we describe the source data and detail efforts to construct a complete
catalogue of stars. In Section \ref{three} we use proper motion survey techniques to obtain a subsample of WD candidates from our
catalogue, and in Section \ref{four} we review the catalogue with emphasis on the completeness and likely mixture of kinematic
populations present. In Section \ref{five} we describe the traditional $\frac{1}{V_{\textrm{max}}}$ approach to measuring the
luminosity function from a proper motion catalogue, and in Section \ref{six} we use this technique to measure luminosity functions
for various velocity subsamples drawn from our WD catalogue. In Section \ref{seven} we present our statistical approach to 
separating the kinematic populations, and measure WDLFs for the thin disk, thick disk and spheroid. In Sections \ref{eight} and 
\ref{nine} we compare our results to those of other studies and draw our conclusions.
\section{The Data}
\label{two}
The SuperCOSMOS Sky Survey (SSS) was compiled by digitizing several generations of photographic Schmidt plate surveys.
The source material includes the second generation POSSII-$B,R,I$ and SERC-$J/EJ,ER$/AAO-$R,I$ surveys, observed by the
Palomar Oschin Schmidt Telescope in the northern hemisphere and the United Kingdom Schmidt Telescope in the south.
These were performed in the photographic $b_J$ $r_{59F}$ $i_{N}$ bands. The subscripts refer to particular combinations of 
filter and photographic emulsion used by the surveys; we ignore the small differences in response between the northern 
and southern hemisphere. 

The first generation POSSI-$E$ and ESO-$R$ surveys provide an early epoch for astrometric constraint. 
The photometry is mostly redundant due to a similar response to $r_{59F}$, but we will nevertheless 
refer to magnitudes in these surveys as $r_{103aE}$ and $r_{63F}$ when distinction has to be made.
Note that $r_{103aE}$ and $r_{63F}$ are identical in response, but we do expect differences in the survey
characteristics due to e.g. a different copying history prior to digitization.
These surveys cross over at $\delta \sim -18^\circ$, thus the sky is divided into three regions of 
common photographic material. This has implications for the charcteristics of the digitized survey. 
Over 1700 fields are required to cover the whole sky, with each field being observed in each of the 
four photographic bands to a depth of $r \sim 21$. The four observations in 
each field are spread over a wide time baseline in the latter half of the twentieth century, with up 
to fifty years between epochs in the extreme.

Schmidt plates were digitized by the SuperCOSMOS scanning machine, scanning at 0.67 arcsecond pixel size and 
15 bit digitization. SuperCOSMOS was operated at the Royal Observatory of Edinburgh - see 
Hambly et al.~(2001a)\nocite{hambly2001a} for an introduction to SuperCOSMOS and brief historical census 
of scanning programs. In some cases, glass or film copies of survey plates were scanned rather than originals. 
This includes the entire northern hemisphere. The use of photographic copies was not thought to significantly degrade 
the quality of the scanned data, however we regard each of the surveys
as having independent characteristics in case any differences arise.
Image Analysis Mode software was used to process the raw pixel data into parameterised object catalogues, which were then merged across
the four epochs to measure proper motion and colours.

The SSS data are housed in a relational database at the Wide Field Astronomy Unit, Royal Observatory of Edinburgh.
The original SSS is now contained within the SuperCOSMOS Science Archive, with
access provided via a web-based SQL query form at \verb"http://surveys.roe.ac.uk/ssa/". Parametrised object 
information is dispensed from two large tables; the \verb"detection" table contains roughly $6.4$ billion 
individual object detections, which are merged into around $1.9$ billion multi-colour, multi-epoch observations 
in the \verb"source" table. A comprehensive description, user-guide and technical notes for the SSS were released 
in a series of three papers - see Hambly et al. 2001a,b,c\nocite{hambly2001b,hambly2001c}.

\subsection{Proper motion limits}
The input data for our proper motion survey is drawn from two sources. The online SSA provides low proper motion
coverage, from a rough lower $5\sigma$ detection limit of $0\farcs05$~yr$^{-1}$ to a fixed upper limit of $0\farcs18$~yr$^{-1}$. The completeness of
the SSA at this upper limit varies widely between fields, due to the differing epoch spread and fixed search radius 
for object pairing. Early tests revealed that fields start to become incomplete above around $0\farcs08$~yr$^{-1}$, as indicated by a 
significant departure from the linear relation $\log(\Sigma N(>\mu)) \propto \log(\mu)$.
Our low proper motion data is thus truncated at an upper limit of  $0\farcs08$~yr$^{-1}$, with lower limits set according to the procedure outlined
in the following section.

Faster moving objects are recovered using a more sophisticated search algorithm described in \cite{hambly2004}. 
Object pairing between epochs works on a field-by-field basis, and starts with the complete
set of parameterized object detections on each of the four plates. Any sources that have been succesfully merged and 
included in the default SSS catalogue at motions lower than the $0\farcs08$~yr$^{-1}$ limit are thrown away.
The remaining objects are then paired in all possible combinations within a search radius set by the chosen upper proper
motion limit of $10\farcs0$~yr$^{-1}$ and the epoch difference between plates. The primary pairing is between the two 
$r$ epochs, which are subject to a magnitude limit of $r_{59F}<19.75$ to reduce noise, but any detections in $b_J$ and 
$i_N$ are folded into the analysis to improve the astrometric solution. To produce an all sky 
catalogue, we take the entire object catalogue and purge multiple observations appearing in plate overlap regions, 
keeping whichever appears closest to its respective field centre in order to match the seaming of the low proper motion 
catalogue. This is far more rigorous than the object pairing in the default SSS, and is designed
to maximise completeness of high proper motion objects at the expense of introducing large numbers of 
spurious detections. Appropriate selection of astrometric and image statistics is necessary to reduce the contamination to
a tolerable level.
\subsubsection{Low proper motion completeness limits}
\label{lowMuData}
\begin{figure}
\centering
\subfigure[]{\includegraphics[width=8cm]{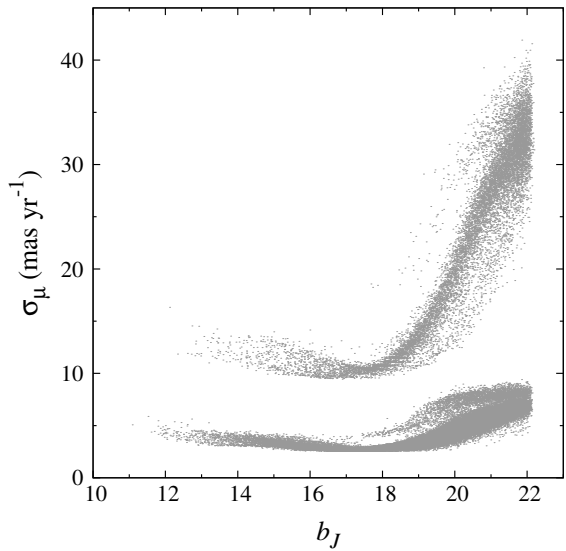}\label{fig:f772lowerMu1}}
\subfigure[]{\includegraphics[width=8cm]{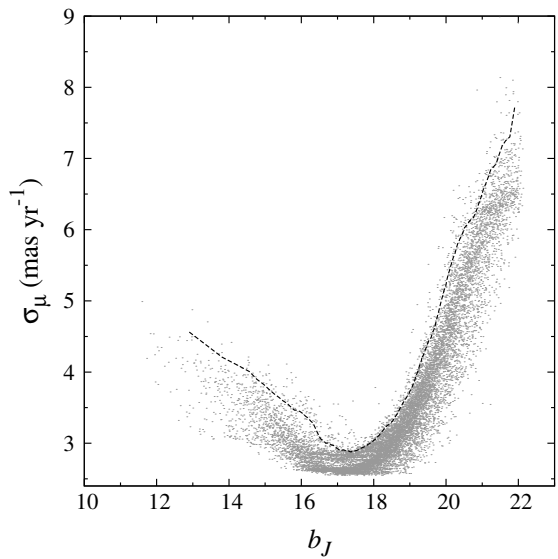}\label{fig:f772lowerMu2}}
\caption[Field 772S proper motion errors]{Field 772S proper motion errors and fitted $\sigma_{\mu}(b_J)$ function. 
(a) includes objects not detected at $r_{63F}$, which form the locus of points at higher $\sigma_{\mu}$. (b) is 
restricted to objects with full four plate detections, and shows the function fitted to the upper boundary of the 
$\sigma_{\mu},b_J$~locus.}
\label{fig:f772sSigmaMu}
\end{figure}

\begin{figure}
\centering
\includegraphics[width=8cm]{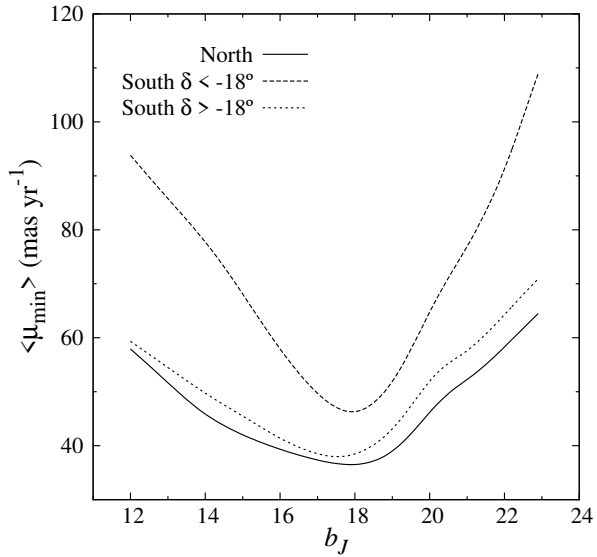}
\caption[Mean lower proper motion limits]{Mean lower proper motion limits in the three sky regions of 
uniform photographic source data.}\label{fig:f772lowerMu3}
\end{figure}

We impose a lower proper motion limit on a field-by-field basis such that all objects have at least a $5\sigma_{\mu}$ proper
motion detection, defined by 
\begin{equation*}
\label{eqn:musig}
\sigma_{\mu} \quad = \quad \sqrt{\left(\frac{\mu_{\alpha}cos(\delta)}{\mu}\right)^2 \sigma_{\mu_{\alpha}cos(\delta)}^2 + 
\left(\frac{\mu_{\delta}}{\mu}\right)^2 \sigma_{\mu_{\delta}}^2}
\end{equation*}
This excludes non-moving objects from our catalogue, and
limits scatter in the reduced proper motion diagram that is used to select white dwarf candidates. 
However, we cannot simply select all objects with $\ge5\sigma_{\mu}$ 
detections, because the proper motion errors show significant scatter at constant magnitude. 
The resulting survey volume limits at given tangential velocity would be unknown, as it would be impossible to 
measure the distance at which the star dropped below the required $5\sigma_{\mu}$ detection threshold. Therefore,
we wish to find the \textit{maximum} proper motion error as a function of magnitude, and use this to fix
the lower proper motion limit. This guarantees that all objects that pass the limit also have $\ge5\sigma_{\mu}$
detections.
The proper motion error $\sigma_{\mu}$ varies from field to field, due to differences in plate quality and time 
baseline. It is also a strong function of magnitude, and at constant magnitude shows a significant spread.
Figure \ref{fig:f772lowerMu1} shows the distribution of $\sigma_{\mu}$ with $b_J$ for a representative field. We use
$b_J$ in this analysis because it is the highest quality photometry available. Objects in 
the upper locus have been missed at one of the four epochs, usually $r_{63F}$, and have inferior astrometric fits. For this
reason we restrict our low proper motion sample to objects with four plate detections.
To fix the proper motion limit, we fit functions to the upper boundaries of the $b_J$,$\sigma_{\mu}$ locus. 
The procedure for each field is as follows. We start at the bright end of the distribution, and bin objects on 
magnitude interactively so that all bins contain one hundred objects. The mean $b_J$ and maximum $\sigma_{\mu}$ in each 
bin is located, after rejecting the top $5\%$ of $\sigma_{\mu}$ as outliers. The full set of 
$\langle b_J\rangle$,$\sigma_{\mu}^{max}$ points defines the
rough upper boundary of the locus, but shows considerable noise on small scales. The raw fit is then processed through 
one stage of smoothing to obtain the desired function. We use a Savitzky-Golay technique to smooth the data, generalized
from their original prescription \citep{savitzky1964} to allow for non-uniform points on the abscissa. This removes small-scale noise while
preserving any low-frequency features, such as the mid-magnitude turning point seen in most fields. Figure 
\ref{fig:f772lowerMu2} shows the same field as (a), but restricted to four plate detections, and with the fitted 
functions shown. The final, smoothed function obtained for each field is denoted $\sigma_{\mu}^{max}(b_J)$, and 
is used to set the lower proper motion limit according to
\begin{equation*}
\label{eqn:musig2*}
\mu_{min}(b_J) = \quad 5 (\sigma_{\mu}^{max}(b_J) + 0\farcs002~yr^{-1}).
\end{equation*}
The set of points defining $\sigma_{\mu}^{max}(b_J)$ is tabulated for each field, and interpolated using
cubic splines to obtain the proper motion limit at arbitrary apparent magnitude. Thus the lower proper motion
limit is a piecewise function of apparent magnitude, which complicates the measurement of the survey volume,
but maximises the sample size by fully exploiting the variable proper motion accuracy. The mean lower proper motion
limit across the survey is shown in Figure \ref{fig:f772lowerMu3}. This is divided into the three sky regions where the 
source photographic data are the same. Note that the lower quality northern data attains a similar
proper motion accuracy as the southern region that shares the same first epoch $r$ material, that of the POSSI-E
survey ($\delta >$-18$^{\circ}$). The ESO-R survey has a much later average epoch, with the result that proper motion measurements are 
considerably more uncertain.
\subsubsection{High proper motion}
\label{highPM}
The precise location of the upper proper motion limit is not as important as the lower proper motion limit from a catalogue
completeness point of view, because of the low probability of finding objects at these velocities and the fact that the
survey volume is not as sensitive to errors at high proper motions ($\sigma_V^2 \propto \mu^{-8}\sigma_{\mu}^2$). However,
we expect to find a greater fraction of elusive spheroid WDs at high proper motions, and extending the range as wide as possible
while limiting contamination from spurious detections allows greater constraint on the luminosity function.

\begin{figure}
\centering
\includegraphics[width=8cm]{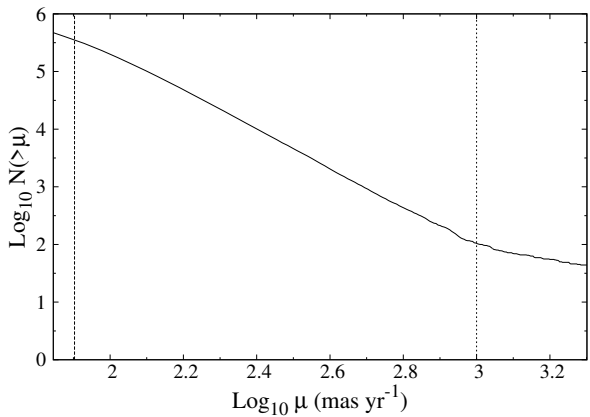}
\caption[Upper proper motion limits]{Completeness limit at high proper motion.}\label{fig:highPMLimit}
\end{figure}

With far fewer high proper motion objects present, there is insufficient statistical power to set an upper limit on a 
field-by-field basis and a global approach is required. We investigated the completeness of proper motions
by looking at the cumulative number counts for all objects, after applying the selection cuts outlined in the remainder 
of this section to eliminate noise. Figure \ref{fig:highPMLimit} shows the cumulative number counts as a function of proper
motion for all survey objects, extending into the low proper motion range to demonstrate the completeness across the interface
as indicated by the first vertical bar. We fix the upper limit at the point where the number counts depart from a straight
line, which was judged to be at $\mu=1\farcs0$~yr$^{-1}$. This is marked by the second vertical bar.
%
%
%
%
%
%

%
\subsubsection{Astrometric residuals}
\label{astrochisquare}
We place restrictions on the astrometric $\chi_{\nu}^2$ in order to exclude objects with a poor fit to their proper motion solution. Appropriate
cuts are obtained iteratively by checking for contamination in the reduced proper motion diagram and the distribution of 
$\frac{V}{V_{\textrm{max}}}$ for the resulting catalogue. A cut of $\chi_{\nu}^2\le2.0$ is applied to the high proper motion data across 
the whole sky, and $\chi_{\nu}^2\le1.5$ in the southern hemisphere for the low proper motion data.
The low proper motion data is adversely affected in the north by the lower quality plate material, and a restrictive cut of 
$\chi_{\nu}^2\le0.8$ is necessary.

The stars ejected from the survey by these cuts can be accounted for later by including an appropriate correction factor
in the LF measurement. In both the high and low proper motion data, the requirement of a detection at all four epochs
results in a four degree of freedom astrometric fit ($\nu=4$). However, the distribution of $\chi_{\nu}^2$ for objects in both the 
SSA and our high proper motion data is significantly tighter than expected.
We have analyzed the astrometric residuals for objects in a selected SSS field in conjunction with synthetic data,
 and established that the tangent-plane position errors used in 
the astrometric fits are overestimated by a factor of around $\sqrt2$, leading to a $\chi_{\nu}^2$ roughly half its true value.
This makes the correction factor hard to determine accurately. Assuming all $\chi_{\nu}^2$ values 
are halved, a $\chi_{\nu}^2\le2.0/1.5/0.8$ cut will eject around $0.5/2/20 \%$ of stars.
%
\subsection{Magnitude limits}
\label{magLimits}
%
%
%
The SSS literature provides magnitude completeness estimates for a selection of $b_J$~and $r_{59F}$~plates in the first
SSS data release, the South Galactic Cap (SGC). These are measured by comparing SSS star and galaxy counts
with those obtained from deeper prime-focus and CCD data in overlapping regions, and estimate near $100\%$
completeness within $\sim1.5$ mags of the plate detection limits.
However, our all-sky survey uses northern Schmidt plate data not included in the SGC, those of the POSS-I and 
POSS-II surveys. Also, the $r_{63F/103aE}$ ~and $i_N$ ~plates are shallower and of lower signal to 
noise than $b_J$ ~and 
$r_{59F}$, and, given the colours of the objects we are interested in, will likely determine the overall 
completeness limits of the SSS fields when considering objects detected on all four plates.
Therefore, it is necessary to estimate new completeness limits for all plates used in the SSS.
Since we cannot obtain deeper imaging for all fields, we instead follow the method of \cite{tinney1993}; 
this involves simulating star and galaxy counts 
along the line of sight, and comparing these to observed counts derived from the corresponding plate material.
We assume that the completeness characteristics of plates within the same photographic
survey are identical, due to uniform quality control, emulsion grade and copying history prior to
digitization. This allows us to restrict our analysis to a representative sample of plates from each of
the eight photographic surveys used in the SSS. Although the plate detection limit varies within each 
survey, we assume that the plates have a common \textit{completeness function}, which we define as the ratio
of detected objects to real objects as a function of magnitude relative to the plate detection limit.
We analyzed five plates from each survey, drawn from five fields in each of the celestial hemispheres.
A summary of the fields selected for analysis is given in Table~\ref{tab:completeness}.
\begin{table}
\begin{center}
\begin{tabular}{cr@{.}lcl}
\hline
Field & \multicolumn{2}{c}{$b$}& $\Omega$ & Surveys \\
\hline                     
\hline                     
411S  & -86&89 & 0.00756 & \multirow{5}{*}{\begin{tabular}{@{\hspace{-0.5cm}}l}SERC-$J/EJ$\\
							     SERC-$R$/AAO-$R$\\
							     SERC-$I$\\
							     ESO-$R$
					   \end{tabular}}\\
350S  & -80&18 & 0.00749 & \\
241S  & -69&38 & 0.00538 & \\
149S  & -60&60 & 0.00591 & \\
237S  & -50&23 & 0.00731 & \\
\hline
507N  & 87&85 & 0.00759 & \multirow{5}{*}{\begin{tabular}{@{\hspace{-0.5cm}}l}POSSII-$B$\\
							    POSSII-$R$\\
							    POSSII-$I$\\
							    POSSI-$E$
					   \end{tabular}}\\
382N  & 80&98 & 0.00751 & \\
270N  & 70&38 & 0.00750 & \\
273N  & 59&44 & 0.00750 & \\
135N  & 49&36 & 0.00723 & \\
\hline
\end{tabular}
\caption[Fields used in completeness analysis]{Fields used to measure the completeness function for each photographic survey.
$b$ and $\Omega$ are the Galactic latitude and solid angle subtended by each field.}
\label{tab:completeness}
\end{center}
\end{table}
\subsubsection{Synthetic star and galaxy counts}
\paragraph{Stars}
\label{starCounts}
Differential star counts along the line of sight to each field are obtained using the Besan\c{c}on Galaxy model 
\citep[see][]{robin2003}. This employs a population synthesis approach to produce a self-consistent model of the 
Galactic stellar populations, which can be `observed' to obtain theoretical data sets useful for testing various
Galactic structure and formation scenarios. We use the coordinates, solid angles and passbands of our selected 
fields as inputs, and select an output number count range that goes several magnitudes fainter than the plate limits.
The SuperCOSMOS filter system is not included in the Besan\c{c}on model; instead we use the CFHT Megacam bands 
$g$ and $r$ to approximate $b_J$ and $r_{59F/63F/103aE}$, and Johnson-Cousins $I$ to approximate $i_N$. 
%
\paragraph{Galaxies}
Within a few magnitudes of the plate limits, galaxies appear as unresolved, point-like objects and have image 
parameters that overlap with stars. We therefore have to include galaxies in our synthetic number counts.
Galaxy counts to faint magnitudes have been determined in many independent studies. We use counts provided
by the Durham Cosmology Group that combine their own results \citep[see e.g.~][]{jones1991,metcalfe1991} with
those of many other authors. These are available online\footnote{See \texttt{http://astro.dur.ac.uk/~nm/pubhtml/counts/counts.html}}
along with transformations to photographic bands. They are provided in terms of log-number counts per square degree per
half-magnitude; we fit straight lines to obtain functional forms for the galaxy counts
in each band, and transform these to 0.1M for comparison to our observed counts. Note that the $\sim25$ square degree
field of view of each Schmidt plate smooths out any anisotropies in the faint galaxy counts.
The fitted functions are given in 
Equations \ref{eqn:galCounts} ($r \equiv r_{59F/63F/103aE}$). The units on the number counts $N$ are
deg$^{-2}$ 0.1M$^{-1}$.
We multiply these functions by the solid angle of each
field, then add them to the star counts to obtain the total synthetic counts for each plate.
\begin{align}
\label{eqn:galCounts}
&\log(N_{b_J}) & = 0.471~b_J &- 7.890 & (16 < b_J < 26)\\
&\log(N_{r})_{\hspace{0.2cm}} & = 0.379~r_{\hspace{0.2cm}} &- 5.351 & (17 < r_{\hspace{0.2cm}} < 25)\nonumber\\
&\log(N_{i_N}) & = 0.606~i_N &- 9.132 & (12 < i_N < 17.75)\nonumber\\
&              & = 0.346~i_N &- 4.397 & (17.75 < i_N < 25)\nonumber
\end{align}
\subsubsection{Observed star and galaxy counts}
\label{obsCounts}
For each field selected for analysis, we obtain object counts to the detection limit on all four plates, binned at
$0.1$ magnitude intervals server-side using an efficient SQL query on the SSA interface. 
We use a wide range of image statistics to include partially resolved galaxies, though close to the plate limits everything is 
pointlike.
\subsubsection{Completeness functions and faint magnitude limits}
\label{maglims}
\begin{figure*}
\begin{minipage}{160mm}
\begin{tabular}{ll}
\includegraphics[width=8cm]{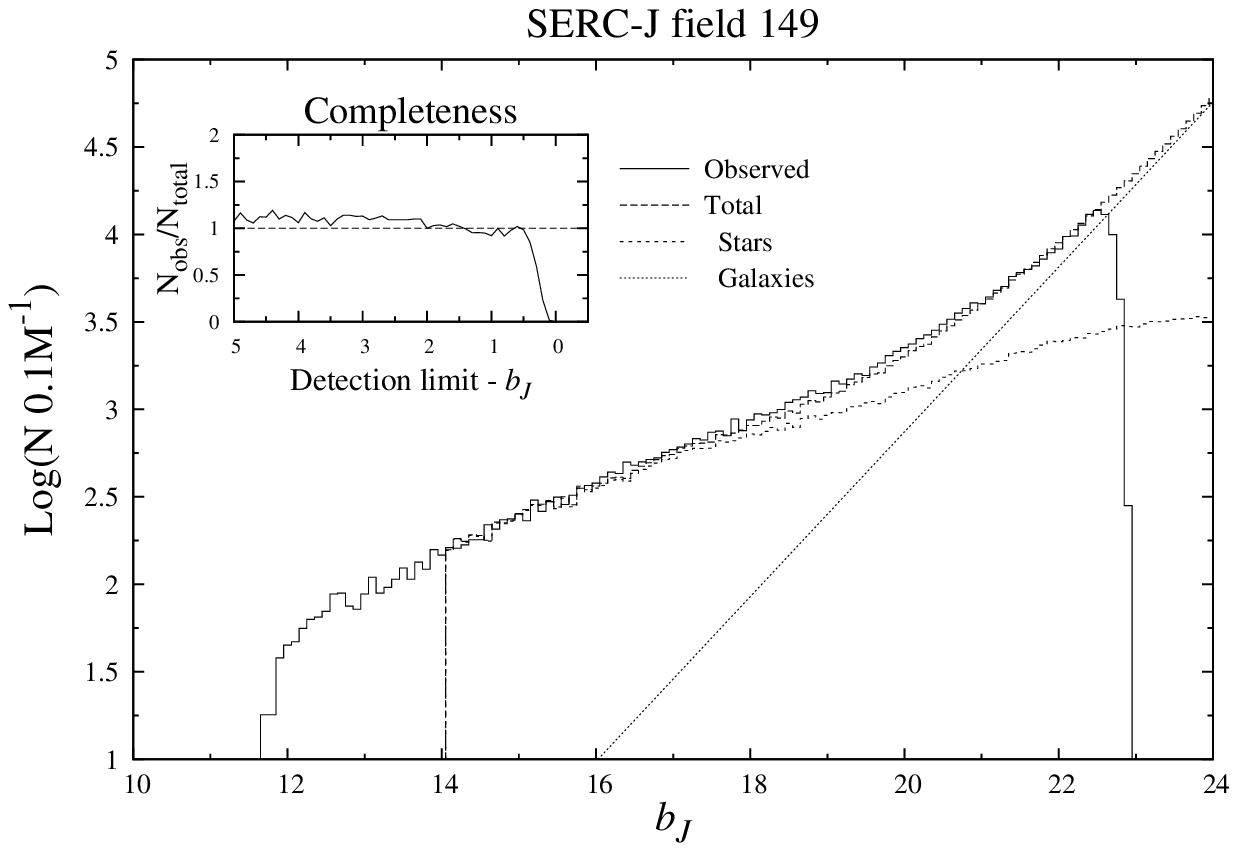}&
\includegraphics[width=8cm]{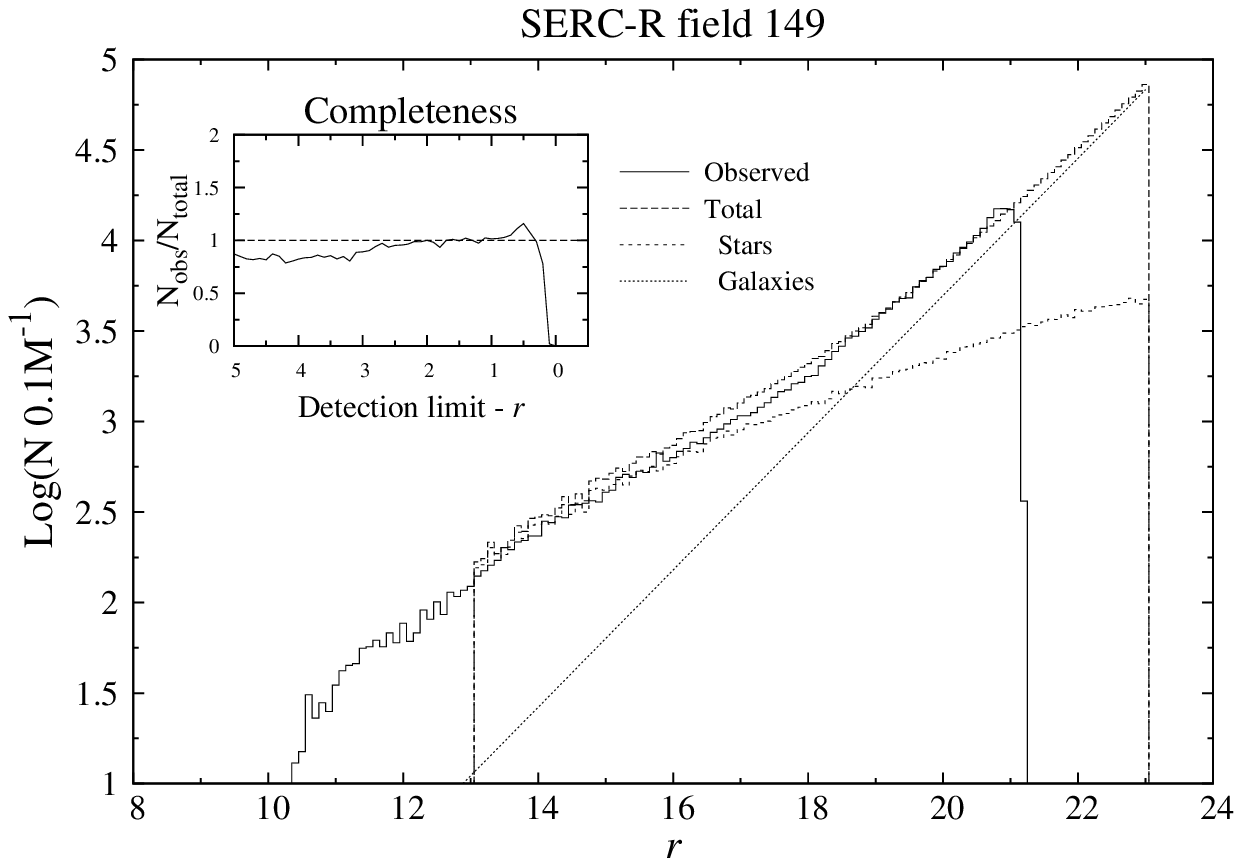}\\
\includegraphics[width=8cm]{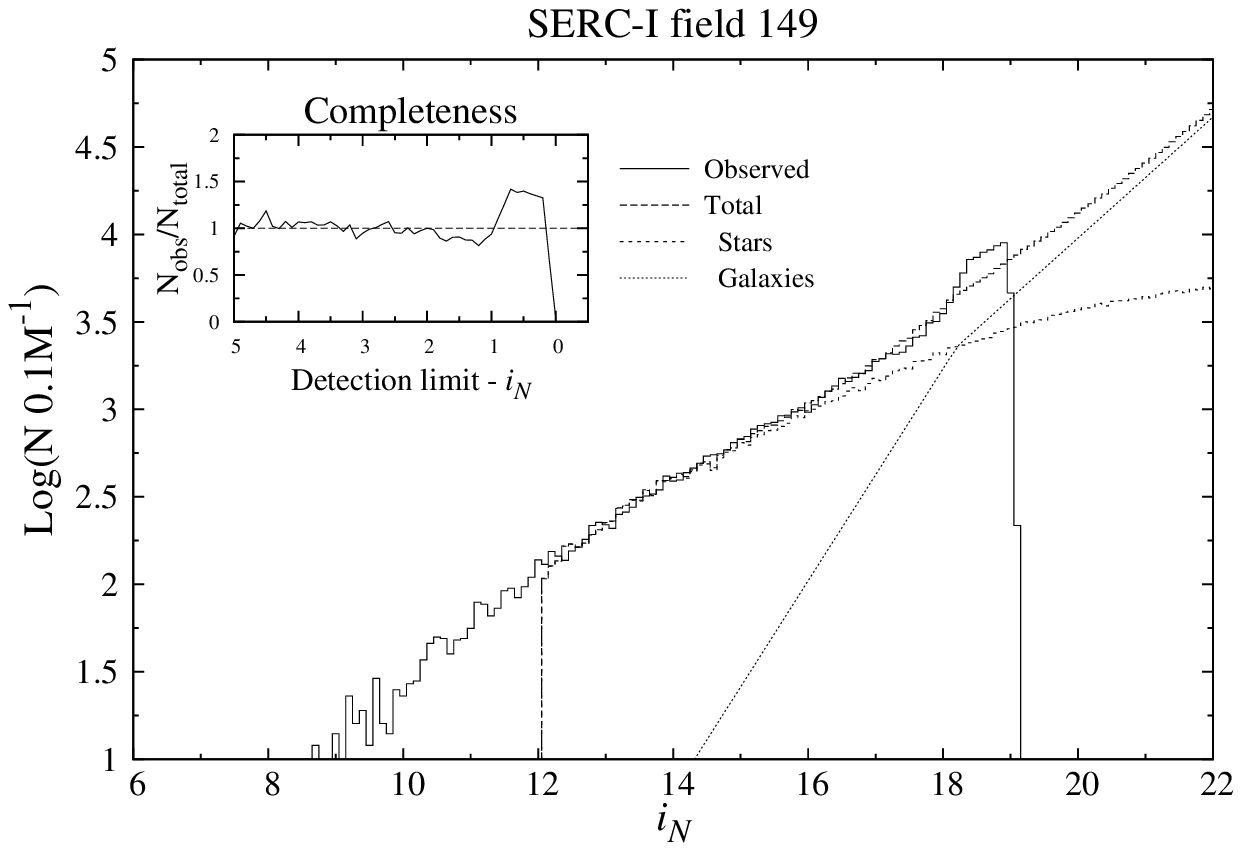}&
\includegraphics[width=8cm]{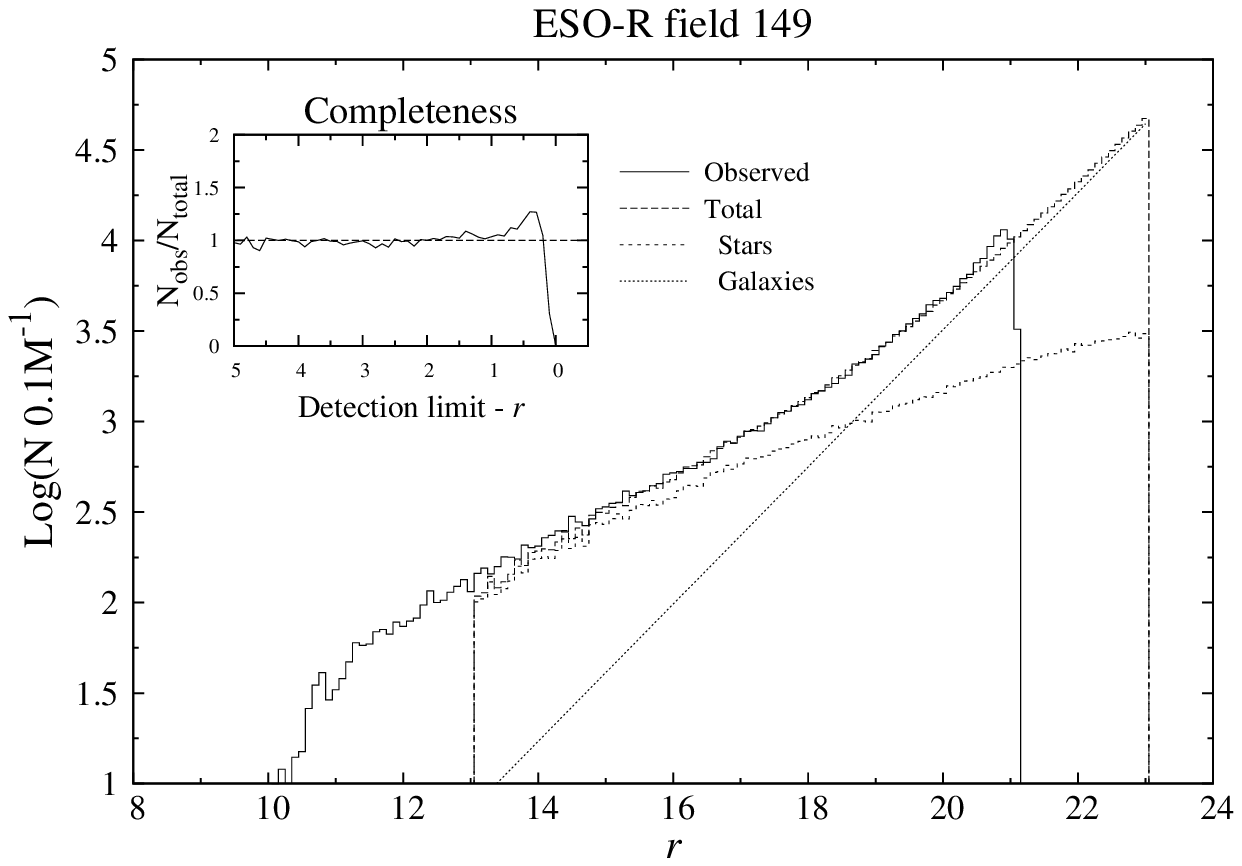}\\
%
\end{tabular}
\caption[Completeness of individual survey fields]{Example completeness analysis for two fields in the SSS, covering all eight of the constituent
photographic surveys.}
\label{fig:compFields}
\end{minipage}
\end{figure*}

\begin{figure*}
\begin{minipage}{160mm}
\begin{tabular}{ll}
\includegraphics[width=8cm]{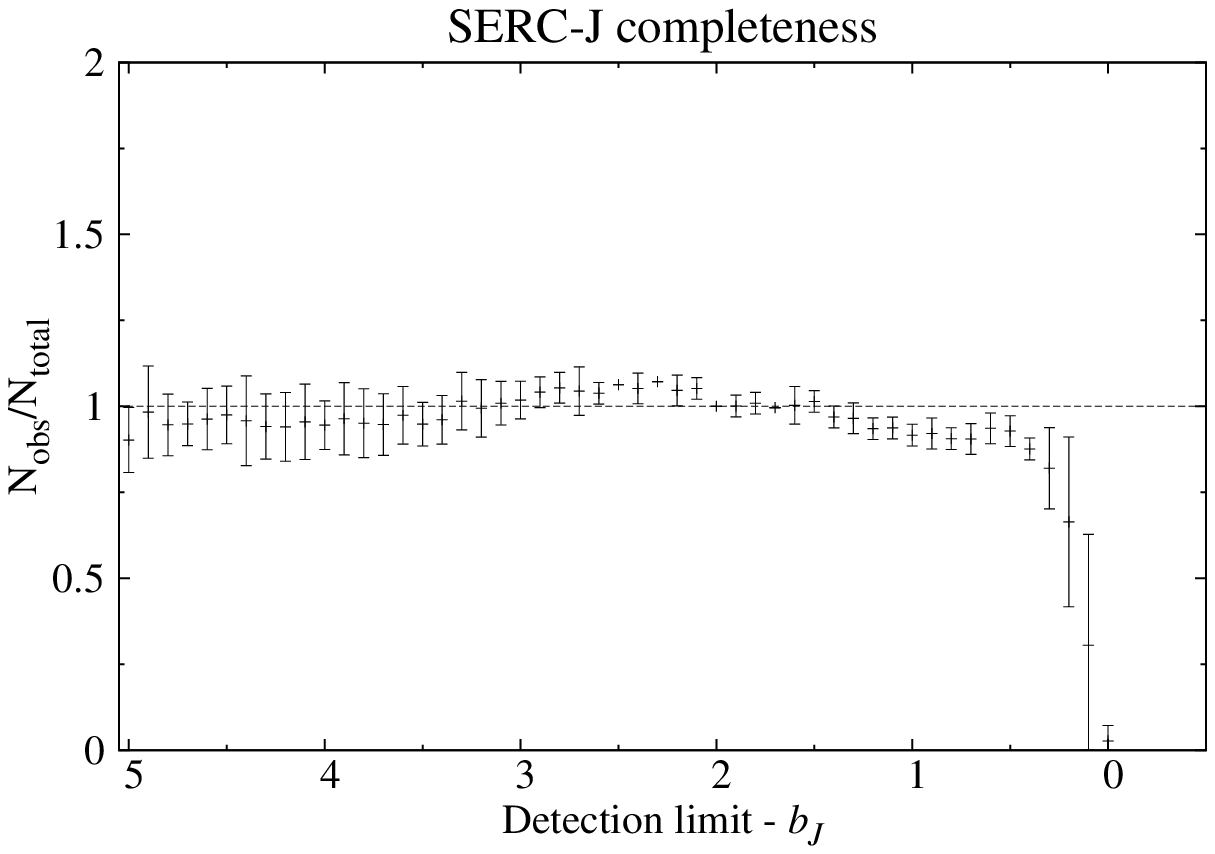}&
\includegraphics[width=8cm]{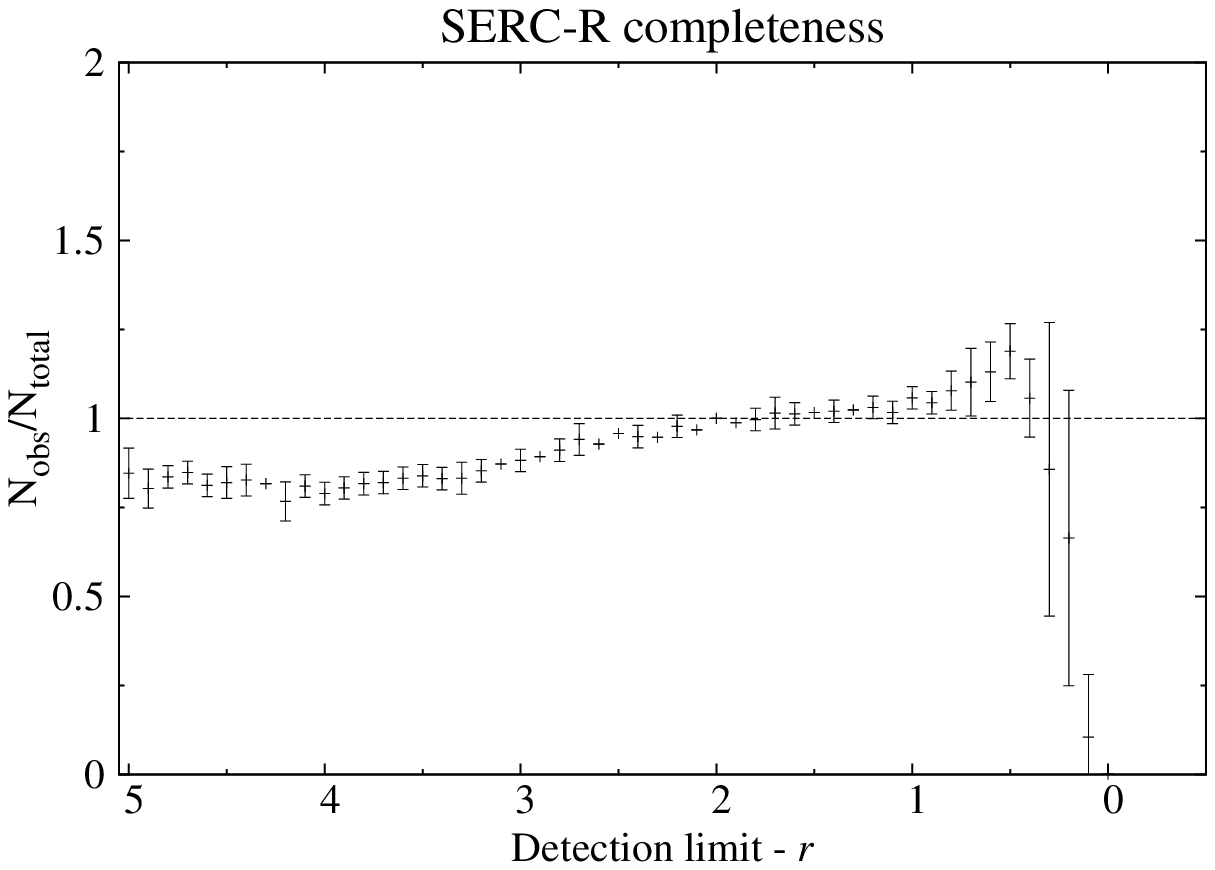}\\
\includegraphics[width=8cm]{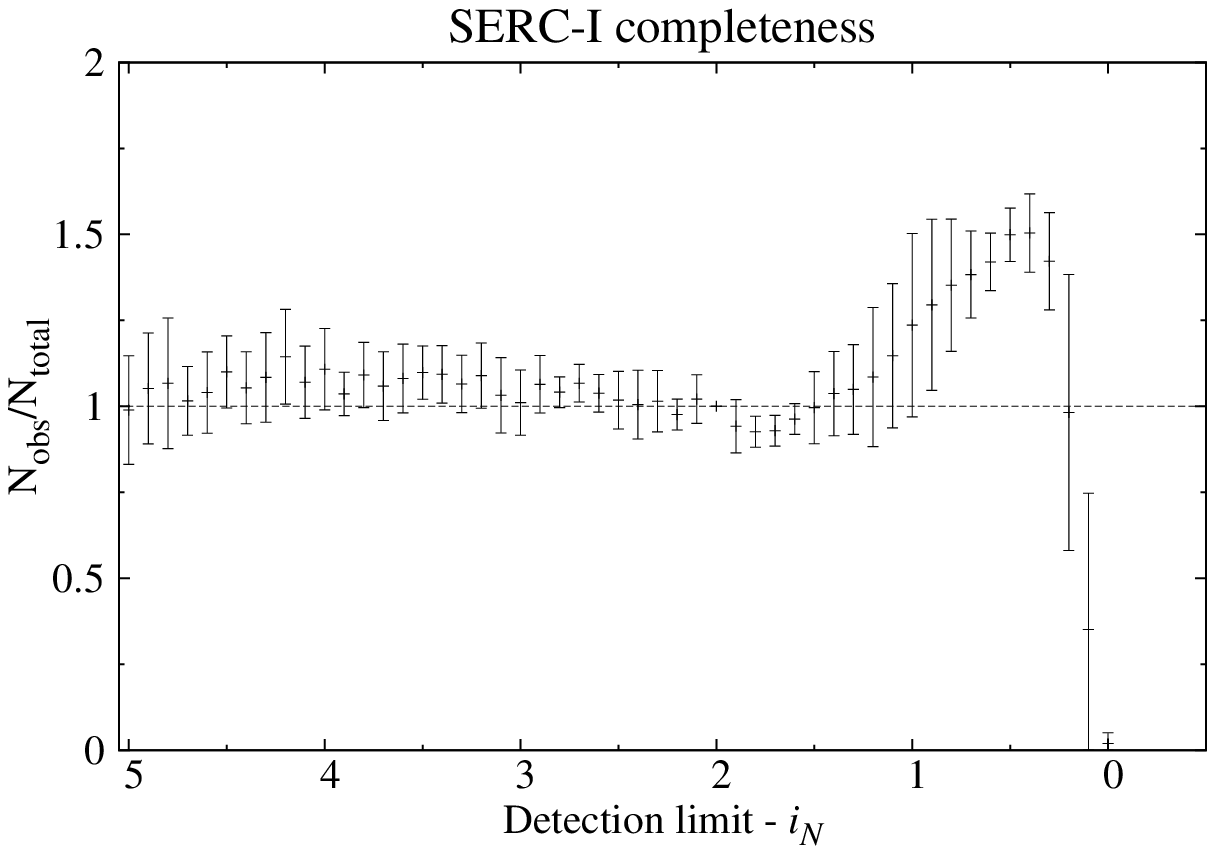}&
\includegraphics[width=8cm]{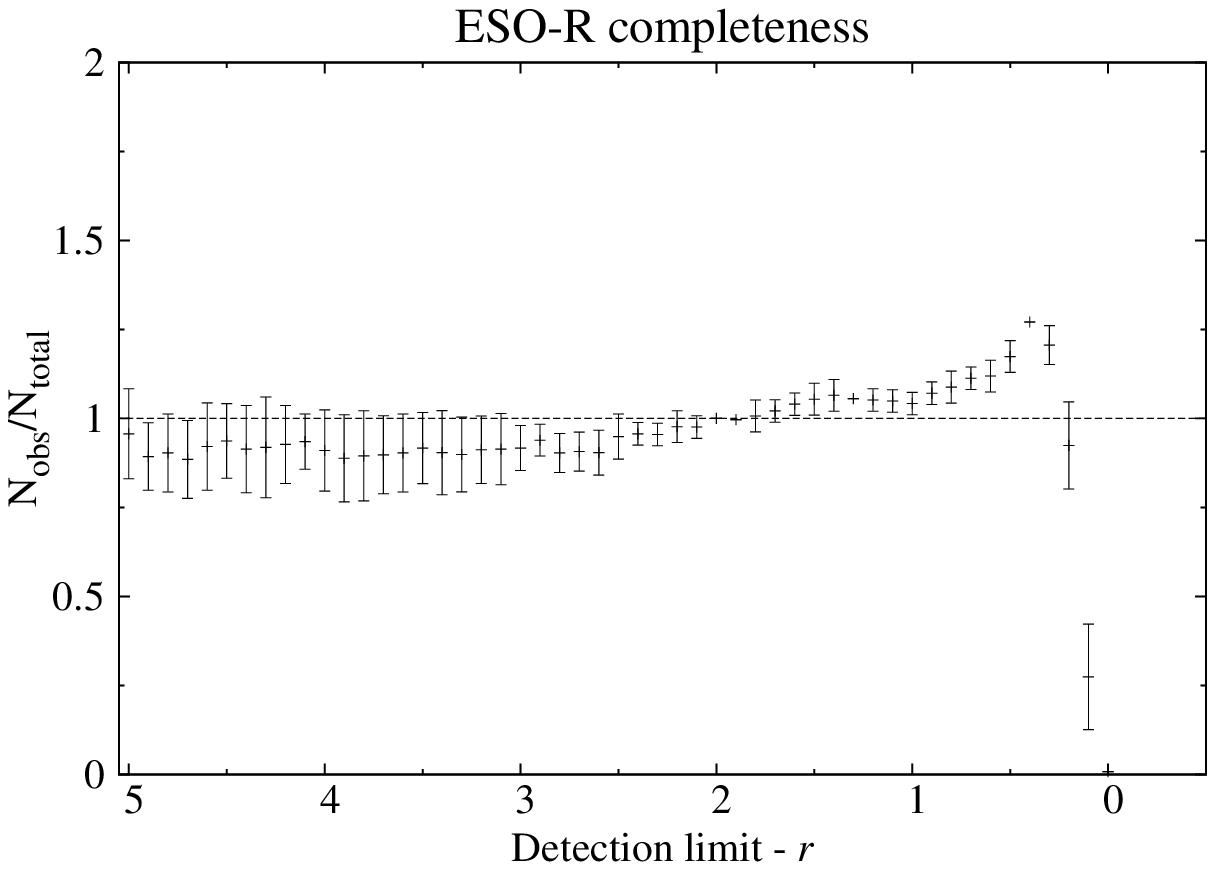}\\
\end{tabular}
\caption[Completeness of photographic surveys]{Completeness functions averaged over five plates in each photographic survey.}
\label{fig:compFuncs}
\end{minipage}
\end{figure*}

In Figure \ref{fig:compFields} we show observed and modelled differential object counts for for each plate in field 270 in the 
north. The ratio of these quantities relative to the detection limit gives the
completeness function for the plate; these are inlaid for comparison. Total model counts are normalised to the observed number 
at two magnitudes above the detection limit, where the plates are assumed $100\%$ complete.
To estimate the global completeness function for each survey, we repeat the analysis for the five representative fields 
listed in Table \ref{tab:completeness} and
take an unweighted average of the individual completeness functions. The global completeness functions
obtained for the northern hemisphere surveys are shown in Figure \ref{fig:compFuncs} for reference.
We set the magnitude limits in terms of an offset from the detection limit, defined as the magnitude of the faintest
detected object. This ensures that the noise at faint magnitudes is avoided.
The global completeness functions are used to judge an appropriate value. 
The offsets selected are listed in Table \ref{tab:maglims}, along with the mean magnitude limit obtained on applying the 
offset to all survey plates.

The superior quality of SERC-$J$ and SERC-$R$ is evident, due to the use of original glass survey plates in the SuperCOSMOS 
scanning program. These plates show $\sim100\%$ completeness to within a few tenths of a magnitude of the detection 
limit. All other surveys were copied photographically at least once before digitizing, which has resulted in noise 
creeping in within a magnitude or so of the detection limit. The POSSII-$I$ survey was copied \textit{twice} before scanning, and the noise
is noticeably worse on these plates. The POSSI-$E$ plates often show extremely large numbers of spurious detections within up to two 
magnitudes of the detection limit. A double peak is often seen in the distribution; this is due to the 
mosaicking of two or more Palomar fields of different depths onto one ESO-SERC field for inclusion in the SSS catalogue.
This means that to ensure completeness, the POSSI-$E$ magnitudes have to be limited to the depth of the shallowest plate 
in mosaicked fields, leading to the large offset of 1.9$m$. This is a significant restriction on the survey volume, considering that
the POSSI-$E$ survey covers the whole sky down to $\delta\sim-18^{\circ}$.
\subsubsection{Astrometric noise at faint magnitudes}
\label{contamination}
Early catalogues drawn using the magnitude limits described in the preceeding Section showed large amounts
of contamination by faint objects with erroneous proper motions. This is most apparent in the reduced proper
motion diagram that is used to select white dwarf candidates (see Section \ref{three}), where large numbers of 
faint disk main sequence stars
with erroneous large proper motions scatter into the region populated by cool white dwarfs. Many attempts were
made to restrict these objects using astrometric and image statistics, but in the end a survey-wide cut on 
the second epoch $r_{59F}$ magnitude was required. We fix this cut at $r\leq19.75$. $r_{59F}$ is used in the
primary multi-epoch object pairing, so is a good proxy for the robustness of the proper motion measurement. The numbers in
parenthesis in Table \ref{tab:maglims} give the mean magnitude limit on combining this cut with the completeness limits.

This involves modelling the effective survey volume over discrete tangential velocity ranges, and turns out to be
a good way to separate the disk and spheroid white dwarf populations, which are distinguishable only on the
basis of kinematic study.
%
%
%
\begin{table}
\begin{center}
\begin{tabular}{lccc}
\hline
Survey&$m_{\textrm{offset}}$& \multicolumn{2}{l}{Mean mag limit}\\
\hline
\hline
\multicolumn{4}{l}{Northern hemisphere:}\\
POSSI-$E$& 1.9 & 18.8 &\\
POSSII-$B$& 0.4 & 22.2 &\\
POSSII-$R$& 0.5 & 20.6 & (19.7)\\
POSSII-$I$& 0.7 & 18.9 &\\
\multicolumn{4}{l}{Southern hemisphere:}\\
ESO-$R$& 0.7 & 20.2 &\\
SERC-$J$& 0.4 & 22.4 &\\
SERC-$R$& 0.3 & 20.7 & (19.7)\\
SERC-$I$& 0.7 & 18.6 &\\
\hline
\end{tabular}
\caption[Magnitude completeness limits]{Offsets defining the magnitude limit for each field in the eight constituent photographic surveys.
Column three gives the average survey magnitude limit on applying these offsets globally. The number in brackets is the corresponding average when the
$r_{59F}<19.75$ constraint is applied to the second epoch $r$ data, as explained in Section \ref{contamination}.}
\label{tab:maglims}
\end{center}
\end{table}
\subsubsection{Bright magnitude limits}
The selection of the bright limits on apparent magnitude is not as crucial, due to the low probability of white dwarfs being
found at these magnitudes. We fix the bright limits at 12 for all bands and across all fields. This is
within the complete range of the Schmidt plates.
\subsection{Photometric accuracy and transmission functions}
While photography cannot compete with CCD astronomy in terms of photometric accuracy, it has been noted in the literature
\citep[e.g.~][]{salim2004} that insofar as digitized photographic surveys go, SSS photometry is of the highest quality with
uncertainties as low as $\sigma_m \sim 0.07$.
Rough estimates of the external error on the photographic \textit{colours} at $b_J\sim16.5$ and $b_J\sim20$ are given in Hambly et al.~2001b 
These are used when fitting photometric models to stellar colours.
We fit a straight line for the error at intermediate and fainter magnitudes, and use the uncertainty at $b_J\sim16.5$ for all brighter magnitudes.
The relation we obtain is
\begin{equation*}
\sigma_{b-r,b-i}(b_J) = 
	\begin{cases}
	0.07& $if $ b_J \le 16.5\\
	0.026  b_J - 0.35& $otherwise.$
	\end{cases}
\end{equation*}
Due to the particular way in which the photometric scale is calibrated, colours are more accurate than single magnitudes.
Uncertainty in single passbands is necessary however for deriving minimum-variance estimates of the photometric distance, by comparing model
and observed magnitudes. These are taken from Table 12 in Hambly et al.~2001b.
Filter transmission functions have been obtained from from \cite{evans1989} ($b_J$) and \cite{bessel1986} ($r_{63F/103aE}$,~$r_{59F}$~and $i_N$). 
\subsection{Image quality criteria}
Every parameterized object detection in the SSA is accompanied by a set of image statistics, some of which we restrict 
in order that stars included in our survey have high quality, stellar images.
\paragraph*{\textit{Blend number}}
We reject objects that have been de-blended at any of the four epochs. The de-blending algorithm attempts to recover individual
object parameters, but is known to be unreliable.
\paragraph*{\textit{Quality number}}
The quality number is a 32-bit integer flag set so that increasingly significant bits indicate increasingly 
compromised situations encountered during image analysis. We restrict this parameter to values less 
than 128, which indicates an image likely to be affected by a step wedge or other label on the photographic plate.
\paragraph*{\textit{Profile classification statistic}}
The profile classification statistic $\eta$ provides a magnitude-independent measure of the `stellarness' of 
each image, by quantifying the deviation of the radial profile slope from that of a mean stellar template. 
$\eta$ is given in terms of a unit Gaussian statistic, and we accept images with $|\eta|\le4\sigma$.
\paragraph*{\textit{Ellipticity}}
Previous studies utilising digitized Schmidt plate data have placed cuts on the ellipticities $e$ of images, in order
to limit contamination from faint galaxies and noise. However, we have found that the ellipticities provided by the SSA
are extremely noisy at intermediate to faint magnitudes, and that any intuitively sensible cut will result in 
a seriously incomplete sample of stars. For example, real stellar images in the $b_J$ band are often assigned $e > 0.5$ 
within two magnitudes of the plate detection limit. We have ignored this parameter altogether.
\subsection{Sky Coverage}
\subsubsection{Individual field areas}
\begin{figure}
\centering
\includegraphics[width=8cm]{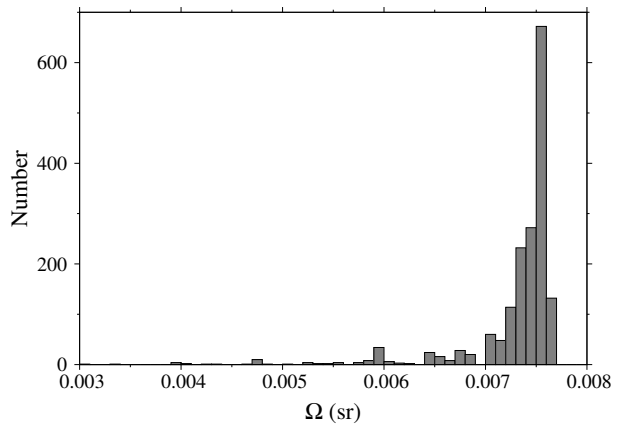}
\caption[SuperCOSMOS field areas]{Frequency of non-overlapping SSS field areas.}\label{fig:areaFrequency}
\end{figure}

The pointings used for the $\sim1700$ fields comprising the SSS use the ESO/SERC system of field centres,
which is based on a $5^\circ$ pitch angle plus some small adjustments to allow for locating
guide stars. This results in $\sim0.5^\circ$ overlap between neighbouring fields, given the 
$\sim 6^\circ \times 6^\circ$ field of view of Schmidt plates. Objects observed multiple times in overlap 
regions are assigned to the field whose centre they are closest to along a Great Circle, providing a `seamless' 
catalogue. The coverage of each field is not given in the SSS database, and has been measured for this work.
This was done numerically
by dividing the sky into small elements of solid angle, accounting for any excluded regions close to the plane, and 
assigning each element to the field it lies closest to along a Great Circle. The average field of view for
fields in the SSS is $\sim0.007$ sr, with a significant spread as shown in Figure \ref{fig:areaFrequency}.
\subsubsection{Bright star haloes and diffraction spikes}
The fraction of each survey plate lost to internal reflections and other spurious images can be significant,
and must be accounted for in order
to accurately calculate the survey volume. The \textit{drilling fraction} is measured exclusively from 
the $b_J$ plates, on which these images have the greatest areal extent,
and ranges from a few percent at high Galactic latitude to more than $50\%$ in the plane.
\subsubsection{Rejection of crowded and dusty fields}
The SSS nominally covers the entire sky. However, in dusty regions at low Galactic latitude the photometric
calibration can be adversely affected by differential extinction across the field. Also, in crowded regions a 
large amount of deblending is required, resulting in significant incompleteness as these stars are 
rejected by our survey procedure, and source merging across multiple epochs suffers from serious confusion.
For these reasons, we avoid the Galactic plane by $10^\circ$ and the Galactic centre by $20^\circ$. We also reject 
53 fields lying slightly outside this region that show a large amount of contamination, manifest in
highly dense regions of stars bounded by field edges. Seven fields centred on the cores of the Magellanic 
clouds are also excluded.
\subsubsection{Rejection of fields with a small epoch spread}
%
%
\label{smallEpochSpread}
In most fields the earliest epoch of observation is $r_{63F/103aE}$, with the remaining three distributed over a
$\sim10$ to $20$ year period. The primary object pairing is between the two $r$ epochs,
and if the $b_J$ and $i_N$ observations are taken very close to the $r_{59F}$ observation, they provide very little
astrometric constraint. 
In such cases, a large number of first-epoch pairings are generated for every real high 
proper motion object. As the search algorithm proceeds through every possible combination within the search radius,
the single correct image pairing is swamped by noise.
Tests show that at proper motions lower than around $0\farcs18$~yr$^{-1}$ this is not a problem, 
as the search radius for image pairing is small.
However, at larger proper motions the effect can be catastrophic.
We therefore reject objects travelling faster than $0\farcs18$~yr$^{-1}$ in any field that has $b_J$,~$r_{59F}$ and $i_N$ taken within 
%
%
1.5 years. 
58 fields fall into this category, and are excluded. This results in slightly different sky coverage depending on the
proper motion range.
\subsubsection{Total sky coverage}
The total survey footprint at low and high proper motions is given in Table \ref{tab:footprint}.
This excludes the Galactic plane and centre regions, accounts for the bright star drilling fraction, and all fields
rejected from each proper motion range.
\begin{table}
\begin{center}
\begin{tabular}{lcc}
\hline
Survey & $\Omega$(sr)&Fraction of whole sky\\
\hline
\hline
$\mu<0\farcs18$~yr$^{-1}$ & 9.11 & 0.72 \\
$\mu>0\farcs18$~yr$^{-1}$ & 8.80 & 0.70 \\
\hline
\end{tabular}
\caption[Survey footprint]{Total sky coverage of the low and high proper motion ranges.}\label{tab:footprint}
\end{center}
\end{table}
\section{Survey Selection Criteria}
\label{three}
\subsection{Reduced proper motion selection}
The proper motions of nearby stars correlate with distance, in the sense that closer objects are more
likely to show large angular velocities. Proper motion can be combined with apparent magnitude to obtain
a statistic called the \textit{reduced proper motion} $H$, which provides a crude estimate of the absolute 
magnitude.
\begin{align}
H_m & = m + 5 \log_{10}\mu + 5\nonumber\\
    & = M + 5 \log_{10}V_T - 3.38 \label{eqn:rpm2}
\end{align}
Although useless for obtaining accurate stellar distances, $H$ is sufficient to distinguish populations of stars
with distinctly different luminosity calibrations or kinematic properties. The classical tool for exploiting this is
the \textit{reduced proper motion diagram} (RPMD), which plots colour against $H$. The RPMD is topologically 
equivalent to the HR diagram, though
with considerable vertical scatter due to the weak correlation between $H$ and $M$. At around ten 
magnitudes fainter than main sequence stars of the same colour, white dwarfs are ideally suited to identification based 
on $H$, and several independent studies \citep[e.g.]{kilic2006} have proved this to be a good way to compile clean white dwarf samples.
\subsubsection{Tangential velocity selection}
\label{vtanselection}
\begin{figure}
\includegraphics[width=8.5cm]{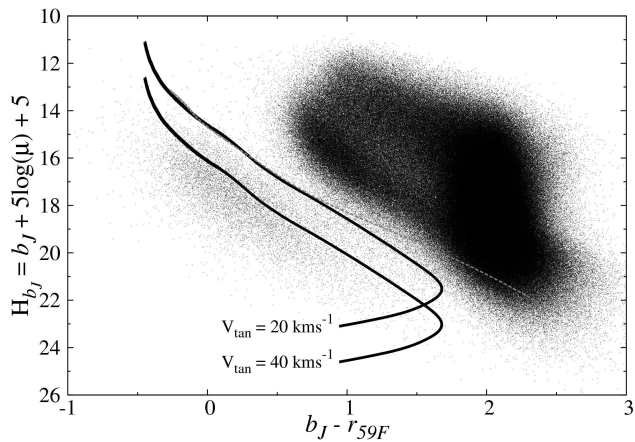}\label{fig:RPMD}
\caption[Reduced proper motion selection]{Reduced proper motion diagram for all stars in the SSA that pass the selection criteria described in Section
\ref{two}. Overplotted are cooling tracks for H (black) atmosphere white dwarfs with $v_{tan} = 20$ and $40$kms$^{-1}$. Also shown is the cooling
track for He (grey) atmosphere white dwarfs with $v_{tan} = 20$kms$^{-1}$. The H tracks are used to identify white dwarf candidates; all objects lying
below a certain line are selected.}
\label{fig:mainRPMD}
\end{figure}

Equation \ref{eqn:rpm2} suggests that with an appropriate colour-magnitude relation, regions of the RPMD inhabited
by white dwarfs of different tangential velocity can be isolated. This allows us to perform rigorous selections on
$H$ to produce catalogues of white dwarf candidates within a well defined tangential velocity range. This is desirable because
cool, low velocity white dwarfs overlap in the RPMD with high velocity subdwarfs from the Galactic 
halo. Contamination by subdwarfs can be reduced by applying
a minimum tangential velocity threshold to white dwarf candidates, producing a cleaner sample of
white dwarfs by restricting selection to regions of the RPMD more widely separated from the subdwarf locus.
Figure \ref{fig:mainRPMD} demonstrates the selection of white dwarf candidates based on reduced proper motion.
 The fact that
low velocity white dwarfs are lost from the survey is of course a drawback of this technique; however, the fraction of stars
that fall below the chosen threshold can be calculated, if the kinematic properties of the population are known.
This is done in each field by projecting the velocity ellipsoid onto the tangent plane, correcting for the mean  
motion relative to the Sun, and marginalising over the position angle to obtain the distribution in tangential velocity
- see \citet{murray1983}.
%
%
%
The values adopted for the mean reflex motions and velocity dispersion tensors are given in Table \ref{tab:velEllip}. 
These are obtained by \citet{chiba2000} for the thick disk and halo; for the thin disk we use the \citet{fuchs2009}
study of SDSS M dwarfs, with values taken from their 0-100pc bin that is least affected by the problems associated
with the deprojection of proper motions away from the plane \citep{mcmillan2009}.
Mean motions are relative to the Sun; the usual Galactic frame is used in which the velocity dispersion tensor
is diagonal in $\sigma_U^2$, $\sigma_V^2$, $\sigma_W^2$. 
%
\begin{table}
\begin{center}
\begin{tabular}{lcccccc}
\hline
Population &$\langle U\rangle$&$\langle V\rangle$&$\langle W\rangle$&$\sigma_{U}$&$\sigma_{V}$&$\sigma_{W}$\\
\hline
\hline
%
Thin disc  & -8.62 & -20.04 & -7.10 & 32.4 & 23.0 & 18.1 \\
Thick disc & -11.0 & -42.0 & -12.0 & 50.0 & 56.0 & 34.0 \\
Halo       & -26 & -199 & -12 & 141.0 & 106.0 & 94.0 \\
\hline
\end{tabular}
\caption[Kinematic properties of the disks and spheroid]{Kinematic quantities adopted in this work. The usual Galactic coordinate axes $UVW$
are used, with $U$ pointing towards the Galactic centre, $V$ in the direction of rotation, and $W$ towards the NGP. The velocity dispersion tensor
is assumed diagonal in this frame. Mean motions are relative to the Sun.}
\label{tab:velEllip}
\end{center}
\end{table}
\subsection{Photometric parallaxes}
Photometric distances are obtained by fitting the two-colour photometry to the white dwarf
model atmospheres and cooling sequences described in \citet{fontaine2001} and updated in \citet{BLR2001}
and references therein\footnote{See \texttt{www.astro.umontreal.ca/$\sim$bergeron/CoolingModels}}.
They were provided in the SuperCOSMOS bands by Dr.~Bergeron on request.
The models consist of cooling sequences for white dwarfs of different surface gravity and H/He atmosphere type.
The gravity and atmosphere effect the fitted distances by changing the absolute magnitude
at a given colour, but with only two data points each we cannot fit these for our stars. 
Instead, we assume $\log g = 8.0$ for all our stars, and fit both hydrogen and helium atmospheres.
It is well known that the gravities of white dwarfs are tightly distributed about this value (for example,
\citet{BLR2001} find $\langle\log g \rangle = 8.070 \pm 0.014$), a consequence
of their narrow mass distribution. Low and high mass white dwarfs exist in roughly equal 
numbers ($\sim10\%$ and $15\%$) \citep{liebert2005}, and fitting to $\log g = 8.0$ models has opposite effects on 
the photometric parallax and luminosity function.

Also, the H/He atmosphere type has very 
little effect on the luminosity above around $6,000K$ ($b_J-r_{59F}\sim0.8$). Below this, the assumption of a H 
atmosphere for a He atmosphere star will cause the absolute magnitude to be considerably overestimated, and the distance
underestimated. In general, optical spectra are useless for distinguishing the atmosphere type, because below around $5000K$
all the absorption lines are washed out. There is therefore some ambiguity over the nature of the coolest white dwarfs in our survey.
\subsubsection{Fitting procedure}
The best fitting $\log g = 8.0$ H and He atmospheres are found in a straightforward manner by variance-weighted least squares,
after interpolating the models at $10K$ intervals. The models corresponding to the upper and lower $1\sigma$ confidence boundaries 
are found by $\chi^2_{1\sigma} = \chi^2_{min} + 1$.
We take no account of redenning, and do not expect it have a significant effect due to the proximity of our stars.
Objects with $\chi^2_{min} > 5$ are rejected from the survey; these are mostly unresolved binaries as explained 
at the end of this section.

\begin{figure}
\centering
\includegraphics[width=8cm]{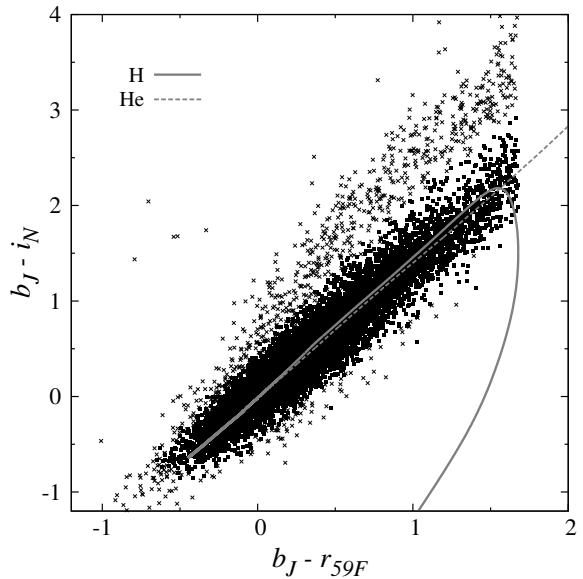}\label{fig:twoCol}
\caption[Two colour diagrams for white dwarf candidates]{Two colour plot for all RPM selected WD candidates. 
Squares indicate stars that pass the $\chi^2<5.0$ cut on the model atmosphere fit; crosses indicate failures.}
\label{fig:twoColour}
\end{figure}

Overall distances are estimated by taking a minimum-variance combination of the estimates from each photometric band
$b_J$, $r_{59F}$ and $i_N$. Note that we avoid $r_{63F/103aE}$ in this calculation. Uncertainty on the overall distance estimate is assigned by averaging
the upper and lower confidence boundaries. A two-colour diagram showing the location and status of our stars relative to the models
is presented in Figure \ref{fig:twoColour}.
%
%
%
%
\subsubsection{Calibration of photometric parallaxes}
\label{calPhotPI}
In order to estimate the accuracy of our photometric parallaxes, we compare the photometric distances
to those obtained by trigonometric parallax for a subset of our white dwarfs. \citet{BLR2001} presented an
analysis of 152 cool white dwarfs with accurate trigonometric parallaxes ($\frac{\sigma_{\pi}}{\pi} < 0.3$), all but two
of which fall within the proper motion and magnitude range of our high proper motion data.
%
%
%
%
%
%
%
Of the remaining 150, 116 have SSS counterparts present in our input catalogues. We investigated the missing stars 
by searching individual plate records
for proper-motion corrected positions; mostly stars are excluded due to lack of a detection at $r_{63F/103aE}$.
Note that when cross identifying stars, we do not apply the magnitude and proper motion completeness limits, 
as the identification is done manually and contamination is not a problem. We do, however, apply the
usual restrictions on blend and quality number, as these directly affect the quality of the photometry. We also require
stars to be detected at all four epochs, which is necessary for the fit. These constraints reduce the sample to 75 stars.

\begin{figure}
\centering
\includegraphics[width=8cm]{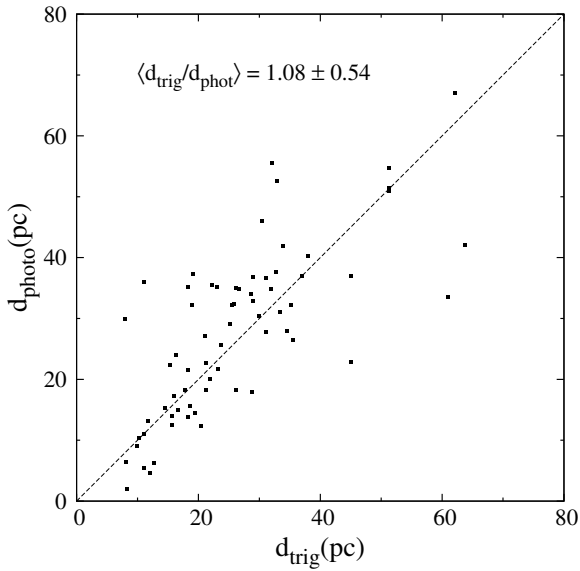}
\caption[Photometric and trigonometric parallaxes]{We compare the distances obtained by our photometric 
parallaxes to those obtained by trigonometric means, for a subsample of stars analysed by \citet{BLR2001}.}
\label{fig:trigVsPhotoPI}
\end{figure}

We fit atmosphere models to the remaining stars, in each case using the appropriate H/He atmosphere as measured
by \citet{BLR2001}. We place the same cut on the residuals as is used in the main survey, in order to remove any stars with
spurious photometry. This results in a sample of 68 stars with adequately fitted photometric distances.
A comparison of the distances obtained by the two methods is presented in Figure \ref{fig:trigVsPhotoPI}. The 
correlation between the two is $r = 0.74$, and $\frac{d_{trig}}{d_{phot}} = 1.08 \pm 0.54$. The error in 
$\frac{d_{true}}{d_{phot}}$ is likely to be lower than this, due to uncertainty in $d_{trig}$, and we estimate the 
accuracy of our photometric parallaxes $\sigma_{d_{phot}}$ to be around $50\%$.

We compared the results of fitting $(b_J-r_{59F},r_{59F}-i_N)$ and $(b_J-r_{59F},b_J-i_N)$ colours to the models. 
The $(b_J-r_{59F},r_{59F}-i_N)$ colours resulted in a slightly worse fit 
($r = 0.76, \frac{d_{trig}}{d_{photo}} = 1.09 \pm 0.55$) 
presumably due to the superior quality of $b_J$, so we adopt $(b_J-r_{59F},b_J-i_N)$ for performing our photometric
parallax fits. 
%
%
%
We also tried relaxing the $\chi^2$ cut to $6$ then $7$; in both cases the sample was 
increased to $70$ stars with $\frac{d_{trig}}{d_{photo}} = 1.08 \pm 0.54$. Therefore, all but a few percent of 
white dwarfs with reliable photometry will pass the survey $\chi^2 < 5.0$ threshold.
\subsubsection{Unresolved binaries}
\label{binaries}
Selection of white dwarf candidates is made on $b_J$,~$r_{59F}$ and $\mu$, so objects with unusual $i_N$ magnitudes turn
up at this point when their $b_J-i_N$ colours are compared to the models. The large populations of objects at red
$b_J-i_N$ in Figure \ref{fig:twoColour} show excess flux in $i_N$, due to an unresolved cool main sequence companion.
These WD+dM binaries are ejected from the survey by the restriction on the photometric parallax residuals, as with at least 
one of three bands polluted no reliable white dwarf fit is possible. This therefore represents a source
of incompleteness in the survey. 
Similarly, Sirius-like systems consisting of a \textit{hot} main sequence star and white dwarf will fail
reduced proper motion selection as white dwarf candidates.

Unresolved double degenerates often have combined spectral energy distributions that closely
resemble single stars of intermediate temperature. They will therefore be successfully fitted to the
models and pass into the white dwarf catalogue as apparently single objects. However, with two stars contributing to 
the flux the derived photometric distance and tangential velocity will be underestimated by up to a factor $\sqrt 2$.

Indications from the local (20pc) white dwarf population are that $\sim70\%$ of white dwarfs exist as single objects
\citep{holberg2009}. Around $20\%$ are members of either a WD+dM or Sirius-like binary system, and so ejection of
these may result in incompleteness of up to $20\%$, depending on what fraction are unresolved by SuperCOSMOS.
Around $10\%$ of white dwarfs exist in double degenerate binaries, which translates to a double degenerate-to-single 
star fraction of up to $7\%$ among our catalogue objects, again depending on what fraction are
not spatially resolved.
\subsubsection{Known ultracool white dwarfs}
\label{ucwdParallaxes}
%
%
Of the dozen or so ultracool ($T_{\textrm{eff}}<4000K$) white dwarfs that have been reported in the literature, 
seven pass the survey constraints and are included in our white dwarf catalogue. These
are LHS 3250 \citep{harris1999,bergeron2002}, CE 51 \citep{ruiz2001}, LHS 1402 \citep{bergeron2005},
SDSSJ0947 \citep{gates2004}, SSSJ1556 \citep{rowell2008}, SDSSJ1452+45 and SDSSJ1632+24 \citep{harris2008}.
Our default photometric parallax method is inappropriate for these stars for two reasons. Firstly, although 
detailed analysis of stars of this class is currently, and necessarily, restricted to only a couple of examples,
it is clear that their properties are quite different to what one would expect based on an extrapolation from higher 
temperatures. In particular, it would appear that most of these objects have extremely He rich atmospheres, which is 
difficult to reconcile with the expected accretion rates of H
from the interstellar medium \citep{bergeron2002}. Also, the single object with a trigonometric parallax (LHS 3250) appears over-luminous 
for its temperature, and has been interpeted either as an unresolved double degenerate or an extremely low mass single 
white dwarf. Secondly, models fail to reproduce the SEDs of these objects for \textit{any} set of parameters, 
indicating incomplete input physics. To quote \citet{harris2008}, writing with reference to their own objects but 
applicable more generally, 
\begin{quote}
``It is premature to add these new 
ultracool white dwarfs to any analysis of the space density and luminosity function of white
dwarfs for two reasons: we do not yet have models to fit the spectra
adequately to give accurate temperatures and H/He abundances, and we do not yet have distances to get luminosities, masses,
and ages."
\end{quote}
While this may be true, these stars are present in our survey and must be dealt with in some way. With these
caveats, we proceed to estimate distances and luminosities for these objects, though note that our adopted values 
should be treated with caution.

Of the seven white dwarfs, only three have anything close to a reliable distance estimate. LHS 3250 has a 
trigonometric parallax as noted above, and CE 51 and LHS 1402 have photometric parallaxes based on spectroscopy 
and multiband photometry, extending into the IR in the latter case. SSSJ1556 has a SED very similar to that of LHS 
3250, and \citet{rowell2008} invoked their similarity to assign a distance by assuming these stars also share 
identical luminosities. \citet{gates2004} performed a similar analysis for
SDSSJ0947 using superior Sloan photometry. We continue in this way for the final two stars. SDSSJ1452+45 is 
closest in colour to LHS3250, though $r_{59F}-i_N$ differs by $\sim0.4$ and this star is most likely warmer. 
SDSSJ1632+24 has identical colours to CE 51 (to $\sim0.01$m),
and we use this star as a reference in this case. The distances and bolometric magnitudes adopted for these 
objects are presented in Table \ref{tab:ucwds}. For all objects, we adopt an identical bolometric magnitude
error of $0.5M$. We have also used SSS proper motions to calculate corresponding 
tangential velocities. Note that on the basis of this distance estimate, SDSSJ0947 
has v$_{tan}=18$kms$^{-1}$ and drops out of the sample. To be clear, it passes the v$_{tan}>20$kms$^{-1}$ 
RPM threshold for the survey, but when the extra (dubious) distance information is added at this stage it 
falls below the cut.
\begin{table}
\begin{center}
\caption[Ultracool white dwarf distances]{Distances and bolometric magnitudes for ultracool white dwarfs 
appearing in our survey.}
\begin{tabular}{llllr}
\hline\label{tab:ucwds}
Star & d (pc) & v$_{tan}$ (kms$^{-1}$)& M$_{bol}$ & T$_{\textrm{eff}}$(K)\\
\hline
\hline
CE 51     & 14.7$^a$ & 44 & 17.5   & 2730   \\
LHS 3250  & 30.3$^b$ & 80 & 16.17  & $<$4000\\
LHS 1402  & 25$^a$   & 58 & 16.8   & 3240   \\
SSSJ1556  & 32$^c$   & 63 & 16.17  & $<$4000\\
J1632+24  & 23$^d$   & 38 & 17.5   & $<$3000\\
SDSSJ0947 & 47$^e$   & 18 & 16.17  & $<$4000\\
J1452+45  & 57$^c$   & 28 & 16.17  & $<$4000\\
\hline
\end{tabular}
\end{center}
\tiny{$^a$Photo $\pi$; $^b$trig $\pi$; $^c$reference to LHS 3250; $^d$reference to CE-51; $^e$reference to LHS 3250 via SDSS colours.}
\end{table}
\subsection{Cool white dwarf atmosphere types}
\label{HHeRatio}
At colours where the choice of atmosphere has a significant effect on the absolute magnitude, a dichotomy arises in the
photometric distance estimate that must be addressed. \citet{knox1999} deal with the unknown atmosphere type 
by assigning half of the stars H atmospheres and half He. However, as He WDs are brighter at a given colour 
they will be sampled over a larger volume and are expected to be present in greater numbers than a simple 50:50 ratio. 
H06
use this fact to estimate the relative numbers of the two types they expect to find in their survey, 
in several bolometric magnitude bins. They also avoid strict atmosphere assignments for each object, choosing instead 
to attach a weight to each atmosphere and allow stars to contribute as both types.

\begin{figure}
\centering
\subfigure[]{\includegraphics[width=7cm]{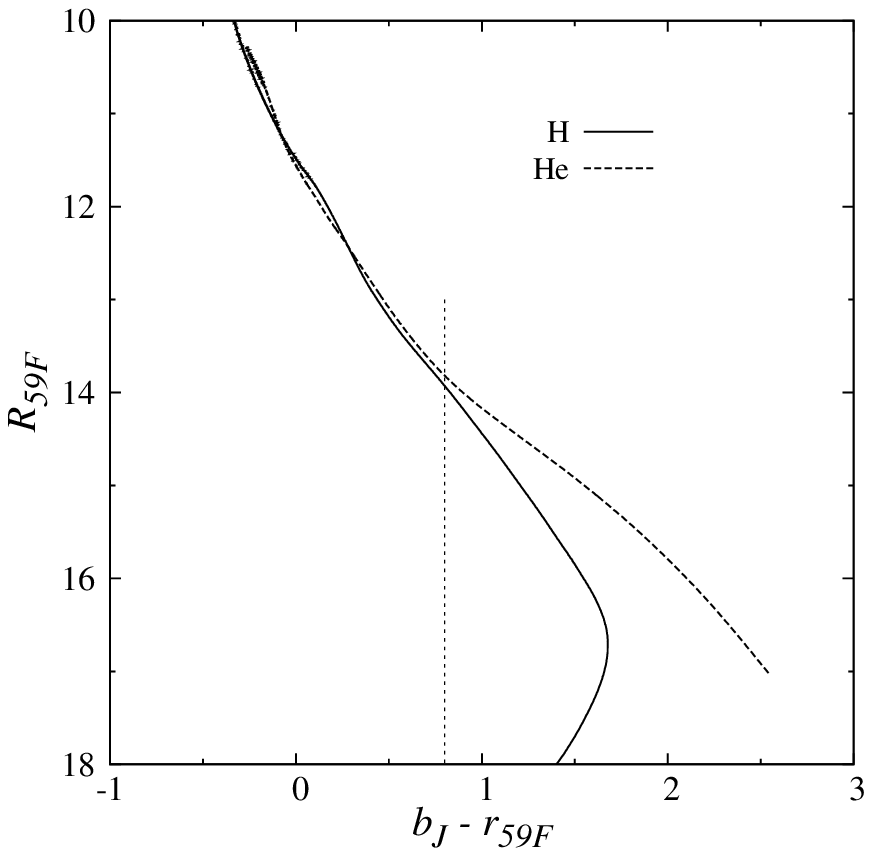}\label{fig:colMag}}
\subfigure[]{\includegraphics[width=7cm]{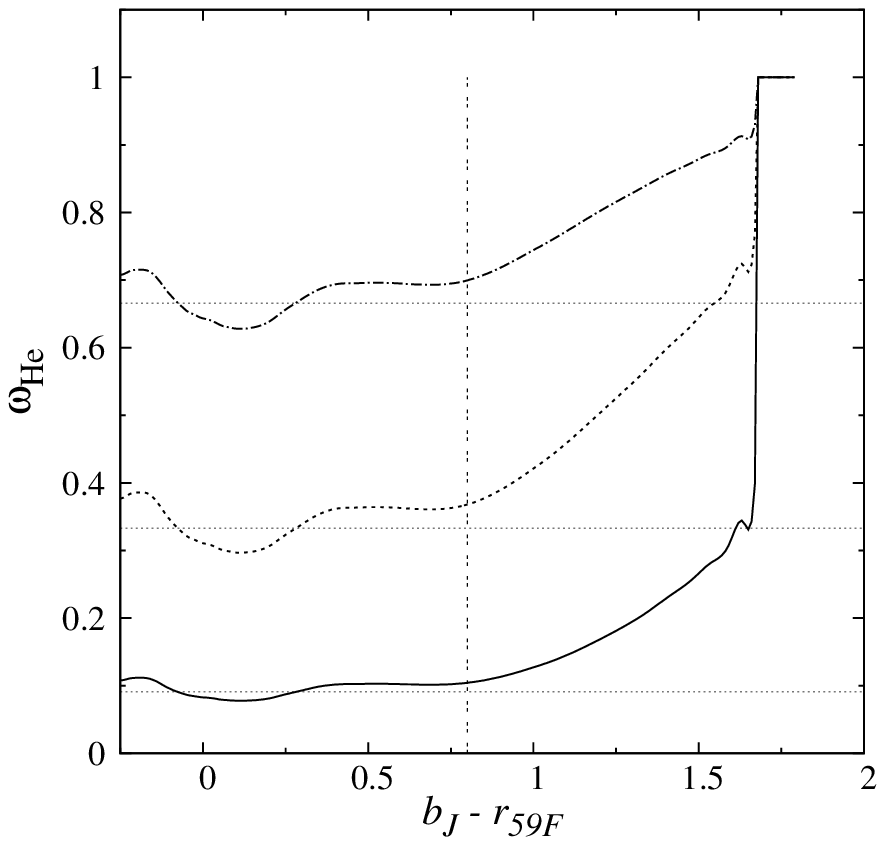}\label{fig:w_He}}
\caption[Helium and hydrogen atmosphere weights for cool WDs]{(a) Colour-magnitude relations for $0.6$M$_{\odot}$ H and 
He white dwarf models, showing the divergence beyond $b_J-r_{59F}\sim0.8$. (b) Weights calculated for cool helium 
atmosphere white dwarfs, under different assumptions for the ratio $\frac{n_{He}}{n_{H}}$ in the solar neighbourhood. 
The curves correspond to $\mathcal{C} = 0.1,0.5,2.0$ (bottom to top), and the straight lines are the weights if the H 
and He colour-magnitude relations did not diverge.}
\end{figure}

We follow the lead of H06, and assign weights to each star depending on colour. However, instead of 
using a few discrete magnitude ranges, we derive a continuous weight function based on the survey volume for each each 
type as a function of colour.
The relative fraction of He to H stars at a given $b_J - r_{59F}$ colour is estimated from the corresponding colour-magnitude
relations $R_{H}(b_J - r_{59F})$ and $R_{He}(b_J - r_{59F})$ assuming a spherical survey volume and uniform
density profile. We obtain the following formulae for the weights $\omega_{H}$ and $\omega_{He}$ for each type;
\begin{align}
\label{eq:hheweight}
\omega_{He}(b_J - r_{59F}) & = \frac{n_{He}}{n_{He} + n_{H}}\\
	    & = \frac{1}{1 + \mathcal{C}^{-1}10^{\frac{3}{5}(R_{He} - R_{H})}}\nonumber\\
\omega_{H}(b_J - r_{59F})  & = 1 - \omega_{He}\nonumber
\end{align}
where $\mathcal{C}$ is the ratio $\frac{n_{He}}{n_{H}}$ of He to H atmosphere white dwarfs in a fixed volume. $\mathcal{C}$ is likely
an evolving function of colour, due to spectral evolution \citep[see e.g.~][]{tremblay2008}. Here we simply 
set it equal to $0.5$, which is the value found by \citet{tremblay2008} for the coolest ($T_{\textrm{eff}}<10,000$K) stars
in their sample. This covers entirely the colour range over which the two types differ in the colour-magnitude plane,
so is an appropriate assumption for all our stars. The helium weight as
a function of colour is plotted in Figure \ref{fig:w_He} for several assumed values of $\mathcal{C}$, along with 
reference weights obtained if H and He white dwarfs did not diverge in absolute magnitude. The diverging colour-magnitude 
relations used to calculate the weights are shown in Figure \ref{fig:colMag}.
%
%
%
%
%
%
%
%
%
%
%
\section{Catalogue summary}
\label{four}
\label{catSum}
We have undertaken a proper motion and magnitude limited survey for white dwarfs in the SuperCOSMOS Sky Survey, using 
two distinct datasets of low and high proper motion stars.
%
%
In both proper motion ranges, most WD candidates are ejected by the $r_{103aE/63F}$ magnitude limit.
While this is undoutedly the largest restriction on the catalogue size, a comparable fraction of low
proper motion WD candidates are ejected by the low proper motion limits.
%
%
In light of Section \ref{calPhotPI}, the catalogue may be up to 50\% incomplete
due to exclusion of blended objects and those for which no first epoch detection exists.
However, in Section \ref{bias} we present evidence that the incompleteness is uniform within
the survey volume and does not bias the catalogue.
%
%
On applying a $v_{tan}>20$kms$^{-1}$ cut in 
reduced proper motion, we obtain
9,749 white dwarf candidates with photometric parallaxes accurate to around 
50\%
%
\footnote{The catalogue is available for download from \texttt{http://surveys.roe.ac.uk/ssa/links.html}}.
Increasing the $v_{tan}$ threshold results in a cleaner catalogue;
$v_{tan}>30$kms$^{-1}$ gives 8206 stars, and $v_{tan}>40$kms$^{-1}$ gives 6592. Note that these numbers are based
on reduced proper motion selected samples; later, we will draw velocity subsamples using the photometric parallaxes
to determine velocities, and the numbers will be slightly different.
A sky projection of all WD candidates that pass the $v_{tan}>20$kms$^{-1}$ cut is shown in Figure 
\ref{fig:lambert}.
\begin{figure}
\centering
\includegraphics[width=8cm]{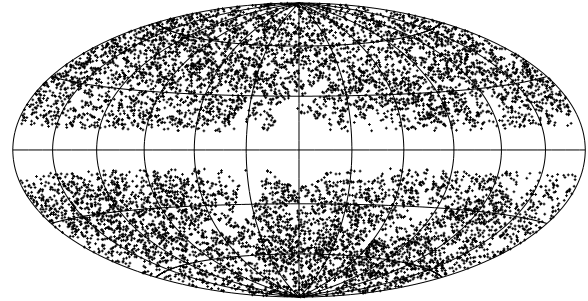}\label{fig:lambert}
\caption[Sky projection of WD candidates]{Hammer-Aitoff projection of WD candidates in the proper motion survey. Galactic coordinates are plotted with $l=b=0$ in the centre.}
\label{fig:lambert}
\end{figure}
\subsection{Survey bias and incompleteness}
\label{bias}
An unbiased survey is uniformly sensitive within the accessible survey volume. The degree to which our survey meets this
requirement can be checked by looking at the distribution of $\frac{V}{V_{max}}$, which follows a uniform distribution
in the ideal case. $V$ is the survey volume contained within the
distance at which the star lies, and $V_{max}$ is the maximum survey volume in which the star could reside and still
be accessible to the survey.
The calculation of the survey volume is non-trivial and will be explained in later Sections;
here we simply present the distribution, in Figure \ref{fig:vvmax}.
The fact the distribution is extremely flat is reassuring; this is evidence that the large incompleteness present
is uniform within the survey volume, and does not bias the survey towards any particular type of star.
As the stellar density profile and velocity ellipsoid are included in the calculation of $\frac{V}{V_{max}}$,
this also suggests that reasonable values have been adopted.
As $\frac{V}{V_{max}}$ is drawn from $U[0,1]$, its distribution should have the property that 
$\langle \frac{V}{V_{max}}\rangle = 0.5 \pm \frac{1}{\sqrt{12N}}$ for $N$ objects. Our $v_{tan} >$ 20 kms$^{-1}$ WD catalogue has
$\langle \frac{V}{V_{max}}\rangle = 0.497 \pm 0.003$.
\begin{figure}
\centering
\includegraphics[width=7cm]{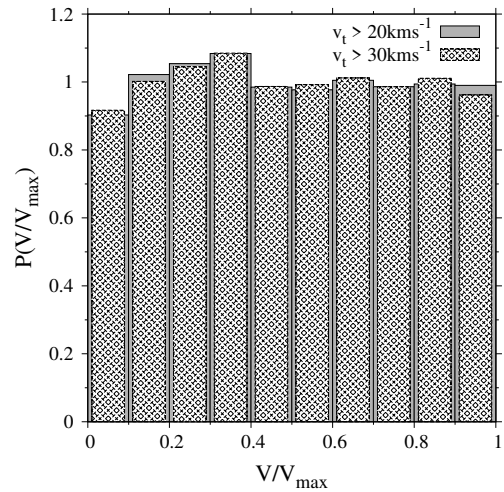}\label{fig:vvmax}
\caption[V/Vmax distribution]{$\frac{V}{V_{max}}$ distribution for WD candidates with $v_{tan} >$ 20 kms$^{-1}$ and 30 kms$^{-1}$. The volume calculation assumes the density profile and velocity ellipsoid of the thin disk for all stars. $\langle \frac{V}{V_{max}} \rangle = 0.497 (0.496) \pm 0.003$ for $v_{tan} >$ 20 (30) kms$^{-1}$, with $\frac{V}{V_{max}}$ following a uniform distribution as expected. Our WD catalogue is consistent with having been drawn from an unbiased sample.}
\label{fig:vvmax}
\end{figure}
\subsection{Distance and tangential velocity distributions}
\begin{figure}
\includegraphics[width=7cm]{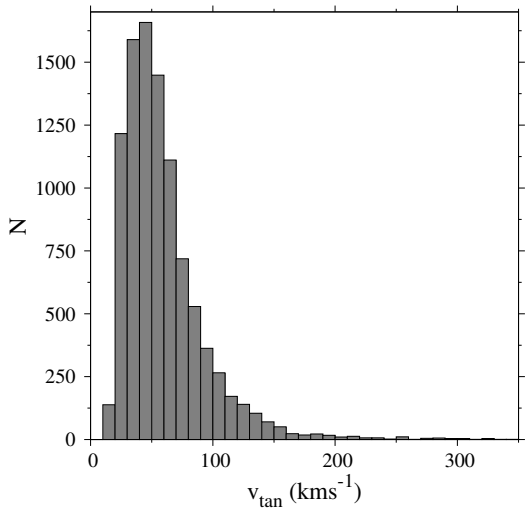}\label{fig:vt}
\caption[Tangential velocity distribution of white dwarfs]{Tangential velocity distribution of white dwarfs in our proper motion survey.}
\label{fig:vtan}
\end{figure}

In Figure \ref{fig:vtan} we present the tangential velocity distribution for all objects in our WD catalogue.
Clearly, a small fraction of stars show velocities much larger than those normally associated
with the thin disk.
Figure \ref{fig:d} shows the distribution of distance for all objects in our catalogue. The vast majority of
stars are within 300pc, suggesting that the lack of a reddening correction does not significantly affect the
photometric parallaxes.
\begin{figure}
\includegraphics[width=7cm]{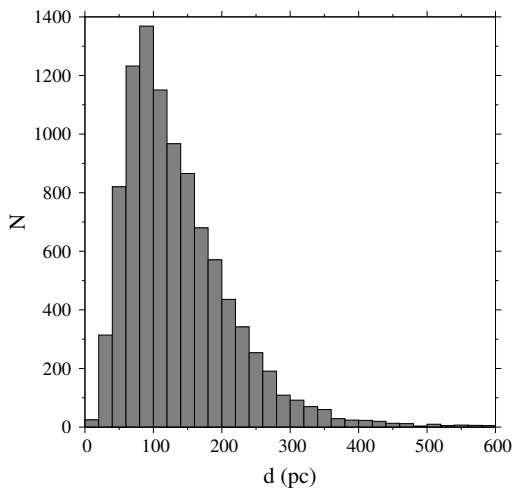}\label{fig:d}
\caption[Distance distributions of white dwarfs]{Distance distribution of white dwarfs in our proper motion survey.}
\label{fig:d}
\end{figure}
%

%
\subsection{Comparison to H06 WD catalogue}
We have compared our WD catalogue to that of H06, which was derived from a combination of
SDSS DR3 and USNO-B data using much the same survey procedure as we have applied here. 1927 of our WD candidates 
with $v_{tan} >$ 30 kms$^{-1}$ lie in the DR3 footprint area, compared to 6009 found by H06. The difference is much
larger than is accounted for by our $\sim50\%$ incompleteness, and is
a result of the wider survey limits used by H06. While our surveys reach a similar magnitude limit (they are after all
based on the same photographic plate material), H06 has a much lower proper motion limit of $\mu \ge 0\farcs02$ yr$^{-1}$.
Our survey is less sensitive to proper motion but covers a much wider area, hence the comparable catalogue size overall.

Of the 1927 stars, 1135 have an unambiguous match in the H06 catalogue. Approximately 800 of our WD candidates do not appear,
presumably due to slight variations in the magnitude completeness limits between USNO-B and the SSS re-reductions of the plate material.
The WD candidates that are common to both surveys show reasonable agreement on stellar parameters: photometric distances
agree to 85\%, and bolometric magnitudes show a dispersion of around 0.5 (assuming hydrogen atmospheres in both cases).
\section{Luminosity Function Methods}
\label{five}
\subsection{The $\mathbf{\frac{1}{V_{max}}}$ density estimator}
There exists a variety of statistical methods for estimating luminosity functions, including both 
parametric and non-parametric methods, maximum likelihood estimators and simple number counts. The 
$\frac{1}{V_{max}}$~technique \citep{schmidt1968} has been used in every major
study of the white dwarf luminosity function, due largely to tradition and to the relative simplicity 
of the approach, although it has the advantage of easily incorporating proper motion selected samples, 
as well as the non-uniform distribution of objects within the Galactic disk. Its performance alongside 
other methods has been analysed by \citet{geijo2006} and found to be satisfactory; 
it provides an unbiased estimate of the true density, and for large enough samples ($\ge300$) accurately 
characterises both the rising slope and faint peak of the luminosity function.

The $\frac{1}{V_{max}}$ method obtains an estimate for the number density of objects $\phi$ by summing 
the inverse of the maximum volume in which each object could reside and still be accessible to the survey,
\begin{equation*}
\phi = \displaystyle\sum_{i=1}^{N} \frac{1}{V_{max,i}}
\end{equation*}
Uncertainties are conventionally assigned assuming Poisson statistics, where the standard error in each 
$\frac{1}{V_{max}}$~term is equal to the term itself (e.g.~$1\pm1$ events). These are then summed in quadrature to 
obtain the error on $\phi$,
\begin{equation*}
\sigma_{\phi}^2 = \displaystyle\sum_{i=1}^{N} \frac{1}{V_{max,i}^2}
\end{equation*}
A more accurate approach would be to use the \citet{gehrels1986} upper and lower confidence 
limits for Poisson statistics, which would result in a $1^{+2.3}_{-0.83}$ contribution from each star for a
$68.27\%$ confidence interval. It is to allow comparison with other studies that we adopt an uncertainty of 
$\pm1$ on the number of stars; all previous measurements of the white dwarf luminosity function have used this.

Objects are binned on bolometric magnitude, and the density associated with each bin is calculated in this manner to obtain
the luminosity function. As in \citet{knox1999}, we plot the luminosity function points at the mean magnitude of the objects
in each bin. This is more realistic, and shifts the observed luminosity function slightly in regions where the number counts
change rapidly with magnitude, such as at the downturn. We also assign horizontal error bars to each point, calculated
by averaging the lower and upper bolometric magnitude errors separately, e.g.
\begin{equation*}
\sigma_{up} = \sqrt{\frac{\Sigma_{i=1}^N \sigma^2_{up,i}}{N}}
\end{equation*}
with $\sigma_{up,i}$~and $\sigma_{low,i}$~assigned from the $1\sigma$~photometric models. 

%
\subsection{Calculating $\mathbf{V_{max}}$}
\label{vmax}
Our survey is limited on both apparent magnitude and proper motion. The intrinsic stellar properties, namely the absolute
magnitude and tangential velocity, lead to restrictions on the distance at which each star could reside and still pass
the survey limits. The apparent magnitude limits are fixed in each field, and the corresponding distance limits are found
according to
\begin{align}
\label{eq:dmaxMag}
d_{max}^m &= \mathrm{min}\left(10^{\frac{m_{i,max} -  M_i }{5}}\right)\\
d_{min}^m &= \mathrm{max}\left(10^{\frac{m_{i,min} -  M_i }{5}}\right)\nonumber
\end{align}
where the index $i$ iterates over each of the four bands.

Due to the non-analytic lower proper motion limits established for each field, no simple expression exists for the corresponding
survey distance limits.
Indeed, if the lower limit changes rapidly with apparent magnitude there may even be several ranges of distance 
in which the star passes the survey limits. $V_{max}$ must be calculated by integrating the appropriate stellar density profile
$\frac{\rho}{\rho_{\odot}}$ along the line of sight between $d_{min}^m$ and $d_{max}^m$, at each step evaluating whether the star passes 
the proper motion limits, which are calculated from the star's apparent $b_J$ magnitude at that distance.
This leads to the integral
\begin{equation*}
V_{max} = \displaystyle\sum_{f=1}^N \Omega_f \int_{r = d_{min}^m}^{d_{max}^m} \frac{\rho}{\rho_{\odot}} \mathcal{P}(r) r^2. dr
\end{equation*}
where the summation is over all surveys fields, and 
\begin{equation*}
\mathcal{P}(r) = 
	\begin{cases}
	1& $if $ \quad \mu_{min}(b_J(r)) \le \frac{v_{t}}{4.74 r} \le \mu_{max}\\
	0& $otherwise.$
	\end{cases}
\end{equation*}
This method for $V_{max}$ follows that of \citet{stobie1989}, generalised to arbitrary
Galactic latitudes by \citet{tinney1993} and further here to allow for the piecewise lower proper motion limits.

\subsubsection{Stellar density profiles}
For the thin and thick disks we use an exponential decay law in Galactic plane 
distance $|z_{\ast}|$ for the stellar density profile, in order to correct for the
truncation of the survey volume by the scaleheight effect. We expect to see no effect arising from the radial scalelength of
the disk, due to the relative proximity of our stars.
The appropriate form for $\frac{\rho}{\rho_{\odot}}$~is thus
\begin{equation*}
\frac{\rho}{\rho_{\odot}} = \exp{\frac{-|z_{\ast}|}{H}}
\end{equation*}
where $H$ is the scaleheight.
We adopt $H=250$pc for the thin disk, which is in line with the result of \citet{mendez1998} 
obtained for faint main sequence stars. These are likely of similar age to the white dwarfs in our catalogue and
are expected to show a similar spatial distribution, having been subjected to the same kinematic heating. This is also
the value used in most other studies of the white dwarf luminosity function, and thus allows more meaningful
comparison with other works.
There is some empirical evidence that the scaleheight of disk white dwarfs increases 
towards fainter magnitudes (see H06) where the stars are on average older,
but we ignore this here. At all but the brightest magnitudes our white dwarfs are so
close to the Sun that the chosen scaleheight makes very little difference anyway.

The Solar distance from the Galactic plane, $z_{\odot}$, is often omitted in studies like this 
\citep[e.g.~][]{tinney1993}, which is equivalent to setting its value to zero. However, the consensus of 
many star count investigations is that in fact $z_{\odot}$ lies close to $\sim 20$pc \citep{reed2006}. 
The adjusted density profile becomes:
\begin{equation*}
\frac{\rho}{\rho_{\odot}} = \exp{\frac{-|r\sin(b) + z_{\odot}|}{H}}
\end{equation*}
where $b,r$ are the Galactic latitude and line of sight distance and $z_{\odot}=20$pc~is the 
Galactic plane distance of the Sun.

For the spheroid, we use a uniform density profile. We expect to see no variation in stellar density over
the distances probed by our survey.
\subsubsection{Corrections}
\label{corrections}
Several steps in the compilation of our white dwarf catalogue have the side effect of excluding a fraction of target stars. 
We correct our density estimate for the ejected objects, under the assumption that the incompleteness is uniform with luminosity 
and does not bias the survey towards any particular type of star.
The discovery fraction $\chi$ of stars that pass the tangential velocity threshold is calculated as described in 
Section \ref{vtanselection}, and the contribution of each star to the total density is adjusted according to
\begin{equation*}
\phi = \displaystyle\sum_{i=1}^{N} \frac{1}{\chi_i} \frac{1}{V_{max,i}}
\end{equation*}
where $\chi$ for each star is taken from the field in which the star was discovered.
A similar correction arises from the restriction on astrometric residuals described in Section \ref{astrochisquare}.
\subsubsection{Atmosphere types}
Fainter than $M_{bol}=14$ ($b_J-r_{59F}\sim0.8$), the choice of H or He atmosphere solution has a significant 
effect on the fitted bolometric magnitude.
In order to account for the unknown H/He atmosphere types of cool white dwarfs, stars are allowed to contribute
as \textit{both} types, with a weight set according to Equation \ref{eq:hheweight}. In each case, $\frac{1}{V_{max}}$
is simply multiplied by the appropriate weight before inclusion in the sum:
\begin{equation*}
\phi = \displaystyle\sum_{i=1}^{N} \frac{1}{\chi_i} \left( \frac{\omega_{H}}{V^H_{max,i}} + \frac{\omega_{He}}{V^{He}_{max,i}}\right)
\end{equation*}
Note that $V_{max}$ is different for
each solution, and they will not in general contribute to the same luminosity bin.
Also, the different photometric distances lead to different tangential velocities, and in some cases only one of the two
solutions will pass the velocity threshold and be included in the LF.

\section{The white dwarf luminosity function}
\label{six}
\begin{figure}
\centering
\subfigure[]{\includegraphics[width=8cm]{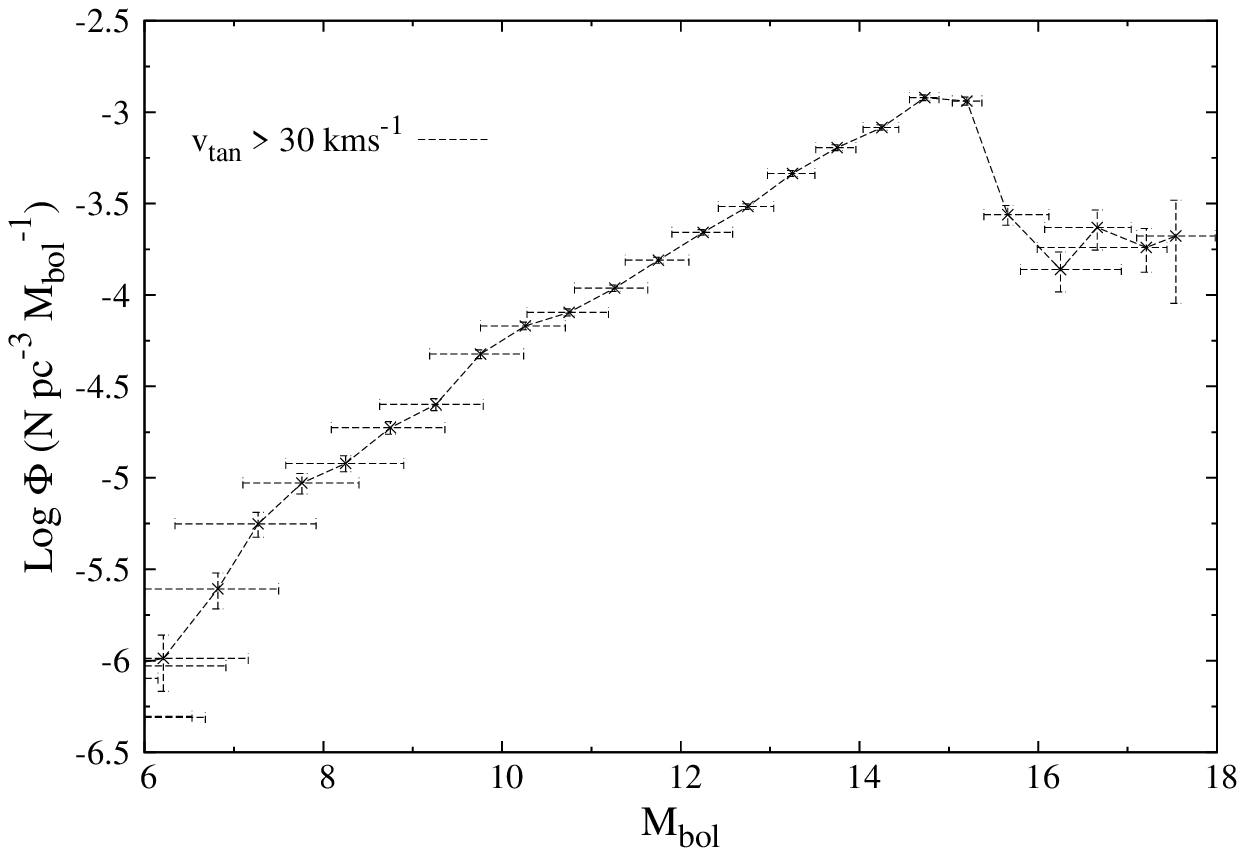}\label{fig:lowvtlf}}
\subfigure[]{\includegraphics[width=8cm]{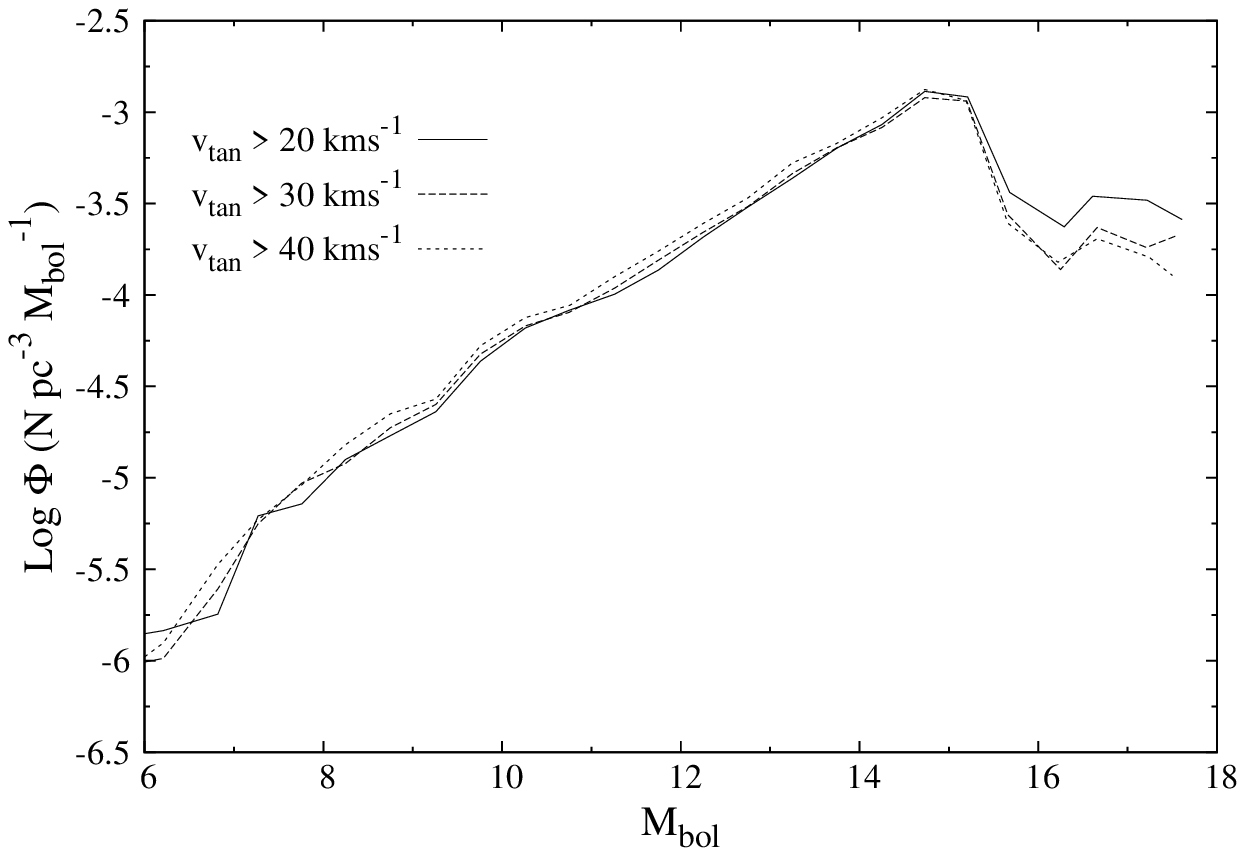}\label{fig:lowvtlfs}}
\caption[The luminosity function for white dwarfs in the SSS]{(a) Luminosity function for 
v$_{t}>30$kms$^{-1}$ white dwarfs in the SSS. (b) Luminosity functions for white
dwarfs on adopting a range of lower tangential velocity cuts.}
\end{figure}

In Figure \ref{fig:lowvtlf} we present the luminosity function for white dwarfs in the
SSS, on adopting a 250pc scaleheight and minimum tangential velocity
threshold of 30kms$^{-1}$. 
The structure in the luminosity function at the faint end is easily discernible - beyond the peak, there is a sharp
drop off followed by a slow decline. Theory predicts that high mass white dwarfs cool faster than their normal
mass counterparts, and, all other things being equal, fall in the region beyond the peak where the luminosity function for
normal mass white dwarfs terminates. Good constraint in this region is vital for obtaining accurate age estimates,
and the number of datapoints beyond the peak is encouraging. However, a quantitative analysis is only possible
in conjunction with theoretical luminosity functions.

Due to the effect of the magnitude-dependent proper motion limits,
at fainter bolometric magnitudes the sample is dominated by stars of brighter apparent magnitude. This improves
the photometric parallax fit and reduces the width of the horizontal error bars on the luminosity function points.
Beyond the peak, the sample is dominated by the ultracool white dwarfs, which have a fixed bolometric magnitude
uncertainty of 0.5$M$ leading to larger error bars relative to the peak.

Figure \ref{fig:lowvtlfs} shows the effect on the luminosity function when the lower tangential velocity threshold
is varied. The fact that no systematic trends are apparent in the luminosity function suggests that the sample is not
significantly contaminated by subdwarfs at low velocities.
\subsection{The luminosity function for high velocity white dwarfs}
\label{spheroidLF}
\begin{figure}
\centering
\includegraphics[width=8cm]{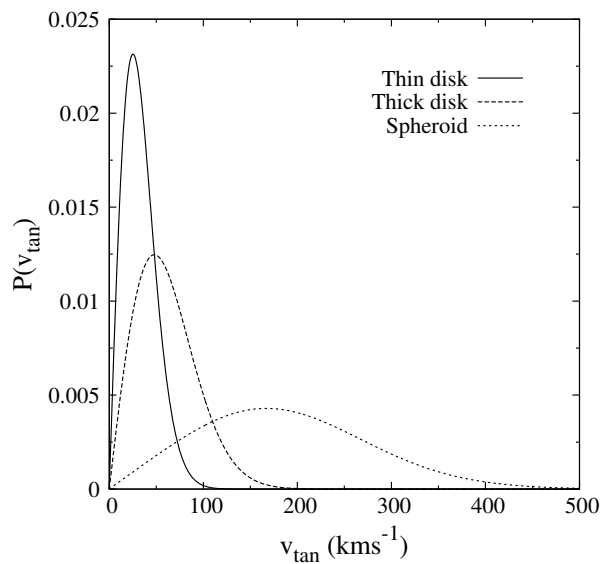}
\caption[Tangential velocity distributions]{Tangential velocity distributions for the three major kinematic 
populations, along the line of sight to field 362 in the southern hemisphere. A cut of v$_{t}>200$kms$^{-1}$
is often considered to cleanly separate the spheroid and disk populations.}
\label{fig:pvt}
\end{figure}

\begin{figure}
\centering
\subfigure[]{\includegraphics[width=8cm]{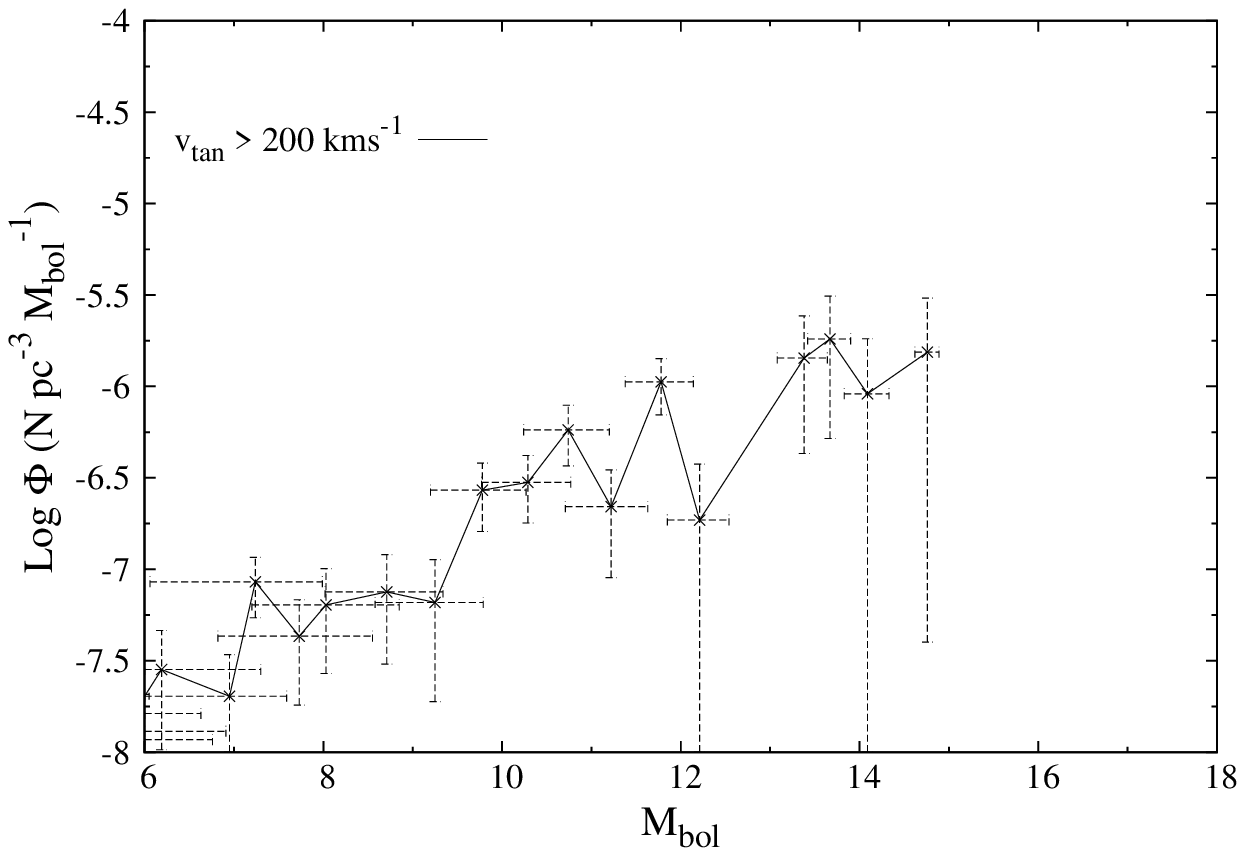}\label{fig:highvtlf}}
\subfigure[]{\includegraphics[width=8cm]{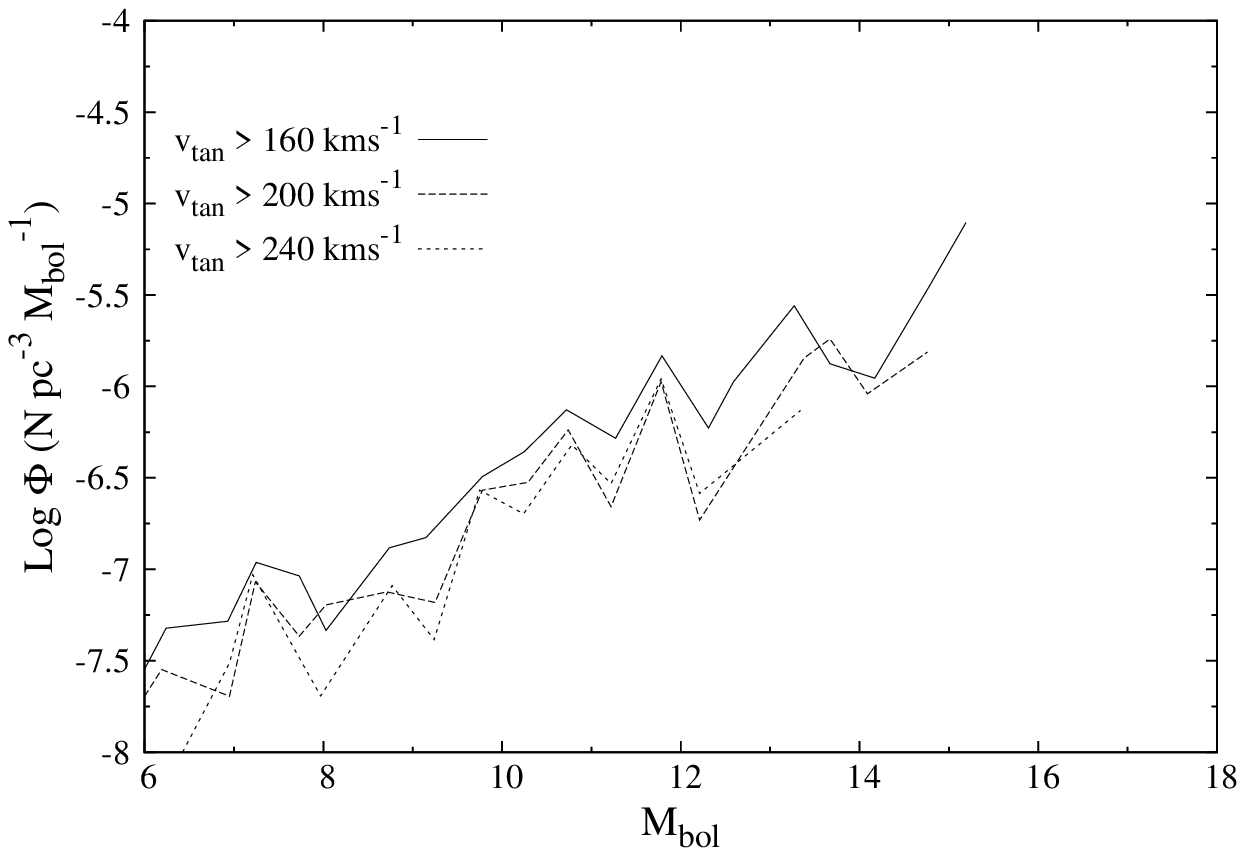}\label{fig:highvtlfs}}
\caption[Luminosity function for high velocity white dwarfs]{Luminosity functions for high tangential 
velocity white dwarfs. Figure (a) shows the LF for v$_{t}>200$kms$^{-1}$ WDs, and figure (b) compares 
the LFs obtained when the lower velocity limit is varied.}
\label{fig:lfHighVelWDs}
\end{figure}

Any spheroid white dwarfs present in our catalogue may be identified by their large tangential velocities.
Figure \ref{fig:pvt} shows the tangential velocity
distributions for the thin disk, thick disk and spheroid along the line of sight to one of our survey fields, as
determined from M dwarfs and low metallicity stars \citep[see][]{fuchs2009,chiba2000}. A cut of v$_{t}>200$kms$^{-1}$
is often considered to cleanly separate the spheroid and disk populations, and the luminosity function obtained
on applying this cut to our catalogue is presented in Figure \ref{fig:highvtlf}. In this case, the discovery fractions 
used to correct for the excluded low velocity stars are calculated from the spheroid velocity ellipsoid,
and the density profile is that of a uniform population. The effect of varying the velocity threshold is
investigated in Figure \ref{fig:highvtlfs}. The fact that the v$_{t}>160$kms$^{-1}$ luminosity function 
sits at a slightly higher density suggests that there is some residual contamination from the disk even at 
these velocities.
%

%
%
The origin of the high velocity white dwarf component can be further probed by looking at the bulk motion.
Attributing individual stars to a particular kinematic population based on tangential velocity alone is 
tricky, because the kinematic properties of cool white dwarfs are relatively uncertain, and it is 
possible that the high velocity tail of the disk population(s) overlaps considerably with the spheroid 
\citep[see][]{reid2005}. 
Ideally,
radial velocities would complete the full 3D space motion in Galactic coordinates, allowing far better 
discrimination for individual stars. 
However, with only proper motions it is still possible to measure the \textit{mean} motion in Galactic coordinates
for the population as a whole, which is sufficient to distinguish a spheroid sample from one drawn from a rotating disk.
This is done by deprojecting the proper motions, according to the method used by \citet{dehnen1998} to analyse
the kinematics of stars in the Hipparcos catalogue.

On doing so, we find that the 93 stars with $v_t>200$kms$^{-1}$ have $\langle UVW\rangle = (-88\pm12,-214\pm24,-38\pm5)$ kms$^{-1}$. 
These would suggest that the high velocity members of our WD catalogue samples are drawn from a non-rotating
population, i.e.~that of the spheroid. We therefore conclude that the luminosity function presented in Figure
\ref{fig:highvtlf} is representative of the spheroid white dwarf population.

\subsection{Local disk and spheroid white dwarf densities}
\label{totalDensity}
Integrating the luminosity function presented in Figure \ref{fig:lowvtlf} gives a total 
local density for white dwarfs in the solar neighbourhood of $(3.19\pm0.09)\times10^{-3}$pc$^{-3}$

The spheroid white dwarf luminosity function of Figure 
\ref{fig:highvtlf} integrates out to $(4.4\pm1.3)\times10^{-6}$pc$^{-3}$, resulting in a 
disk-to-spheroid ratio of $725 \pm 215$ disk stars to every spheroid star.
%
%
%
%
These densities are significantly lower than those obtained by other studies, e.g. H06
find $4.6\times10^{-3}$pc$^{-3}$ and \citet{legget1998} find $3.4\times10^{-3}$pc$^{-3}$ for the local
density of disk white dwarfs, and H06
find $4\times10^{-5}$pc$^{-3}$ for those of the 
spheroid. If we recall from Sections \ref{calPhotPI} and \ref{bias}, our catalogues may be up to
50\% incomplete due to blended objects and those missed at first epoch $r$, which explains the
disagreement between these numbers. This incompleteness is expected to be uniform with bolometric magnitude,
and therefore will not affect the disk-to-spheroid ratio, nor will it affect any conclusions
about the age of these two populations, as this is insensitive to the normalisation of the luminosity function.
It should be noted at this point that while the disk WDLF almost certainly falls rapidly to zero beyond the
magnitude range of our survey, the spheroid WDLF most likely continues rising to fainter magnitudes.
Deeper surveys may find significantly larger integrated densities for the spheroid by including fainter magnitude bins,
and the numbers presented here should be regarded as lower limits.
The contribution from fainter objects may also be constrained using theoretical luminosity functions, however we do not do this here.
\section{Untangling the disks and spheroid}
\label{seven}
\label{untangle}
The technique of drawing velocity selected sub-samples of stars works reasonably well at isolating a 
clean sample of spheroid stars. Although uncontaminated by disk stars, samples drawn in this manner
do of course miss a large fraction of spheroid stars at lower velocities, where they are indistinguishable from
the disk. Also, the thin and thick disks could never be separated in this manner, due to the large overlap in 
tangential velocity leaving no range in which the populations can be reliably isolated from each other
or the spheroid. It is worth pointing out that chemical tagging, as might be used to distinguish the main sequence
members of these populations, cannot be used for white dwarfs as any photospheric metals sink rapidly below
the photosphere.

An alternative approach can be derived by considering the how the sampled volume for each of the 
kinematic populations affects the numbers of stars that make their way into the survey. The total number of
survey stars, $N_{\ast}$, is determined by the local number density $n$ multiplied by the survey volume $V$, 
separately for each kinematic population,
\begin{equation}
\label{nStars}
N_{\ast} = n_{\mathrm{thin}} \times V_{\mathrm{thin}} + n_{\mathrm{thick}} \times V_{\mathrm{thick}} + n_{\mathrm{sph.}} \times V_{\mathrm{sph.}}
\end{equation}
where the subscripts refer to each of the major kinematic populations.
In a narrow range of magnitude, such as one of our luminosity function bins, the factor that determines $V$ for each
of the populations is the tangential velocity distribution combined with the survey proper motion and 
tangential velocity limits. The magnitude limits are effectively decoupled from the analysis, because stars belonging
to each population have very similar mean absolute magnitudes over the small ranges considered. The kinematics differ 
considerably however, and for equal survey limits each population will be sampled over a different volume of space.
By varying the tangential velocity limit, and recalculating each $V$ for the new $N_{\ast}$, a 
set of linear equations in the unknowns densities $n$ can be generated. In general, the 
equation set is non-singular and solvable using linear algebra techniques.

The power of this approach lies in the fact that individual stars are not assigned conclusively to either population. 
Instead, it simply measures the fraction of stars that belong to each population, as a function of $v_t$.
Note that we will refer to the survey volume measured in this technique as the \textit{effective volume}, $V_{\mathrm{eff}}$,
to distinguish it from that used in the conventional $\frac{1}{V_{max}}$ technique.
In this approach no strict distance boundaries arise from the proper motion limits and the survey is sensitive to some 
velocity sub-sample at all distances.
\subsection{Modelling the effective volume}
The survey volumes used in this analysis are calculated in a rather different manner to $V_{max}$ from Section
\ref{vmax}. This is because neither the absolute magnitude nor tangential velocity are directly observed from
the survey objects. In the first instance, we instead use the theoretical mean absolute magnitude for all stars 
in each luminosity bin, obtained by integrating model $M(M_{bol})$ relations for
each survey band. We assume the LF is flat over the width of each bin, and use standard $\log g = 8.0$ DA models;
the difference between the effective volumes for DA and DB models is very small at fixed bolometric magnitude. 
These are then used in conjunction with the apparent magnitude limits in each survey field to place limits on 
the distance at which this hypothetical star could lie and still pass the magnitude limits. These are found
according to
\begin{align*}
d_{max}^m &= \mathrm{min}\left(10^{\frac{m_{i,max} - \langle M_i \rangle}{5}}\right)\\
d_{min}^m &= \mathrm{max}\left(10^{\frac{m_{i,min} - \langle M_i \rangle}{5}}\right)
\end{align*}
which is similar to the expression in Equation \ref{eq:dmaxMag} except the observed absolute magnitude has been replaced
by the theoretical mean for the LF bin.

The effective survey volume for each kinematic population is then found by integrating the appropriate density
profile $\frac{\rho}{\rho_{\odot}}$ between $d_{min}^m$ and $d_{max}^m$, at each step correcting for the fraction 
of stars that pass the tangential velocity limits at that distance. The survey volume is thus marginalised over the
tangential velocity, leading to the effective volume probed for each population.
This is done as follows,
\begin{equation*}
V_{\mathrm{eff}} = \displaystyle\sum_{f=1}^N \Omega\int_{r = d_{min}^m}^{d_{max}^m} \frac{\rho}{\rho_{\odot}} r^2 \chi(r) . dr
\end{equation*}
where the summation is over all survey fields.
The factor $\chi(r)$ is the discovery fraction of stars that pass the tangential velocity limits at distance $r$, and is 
calculated from the cumulative distribution by 
\begin{equation*}
\chi(r) = \textrm{cdf}((v_{upper}(r)) - \textrm{cdf}((v_{lower}(r))
\end{equation*}
where $v_{upper}$ and $v_{lower}$ are tangential velocity limits fixed by the survey design.
The appropriate $v_{upper}$ and $v_{lower}$ are found according to 
\begin{align*}
v_{upper}(r) &= \mathrm{min}(v_{max},4.74 \, \mu_{max} \, r)\\
v_{lower}(r) &= \mathrm{max}(v_{min},4.74 \, \mu_{min} \, r)
\end{align*}
where $v_{max}$ and $v_{min}$ are external tangential velocity limits that may be applied to restrict the
velocity range, and the second 
argument in each case is the velocity limit arising from the survey proper motion limits at distance $r$.

The velocity ellipsoids adopted are those of Table \ref{tab:velEllip}. As before, a uniform density profile
is used for the spheroid and a 250pc scaleheight exponential disk is used for the thin disk. The thick disk 
density profile is that of a 1500pc scaleheight exponential disk; however, the proximity of our stars
means the effective volume is rather insensitive to the scaleheight at this level, and it is the 
velocity correction alone that separates the thick disk and spheroid populations in our survey.
\subsection{Solution for the number density}
\label{NNLS}
Applying external $v_{min}$ and $v_{max}$ cuts enables one to generate several instances of 
Equation \ref{nStars}, which can be used to construct a linear equation set in the unknown densities.
We use non-overlapping $v_{tan}$ ranges to construct each equation, so that the covariance of neighbouring
star count bins is zero.
The set of equations can be cast in matrix form like
\begin{eqnarray*}
\begin{bmatrix}
V_{thin}(\delta v_{1}) & V_{thick}(\delta v_{1}) & V_{sph.}(\delta v_{1})\\
V_{thin}(\delta v_{2}) & V_{thick}(\delta v_{2}) & V_{sph.}(\delta v_{2})\\
V_{thin}(\delta v_{3}) & V_{thick}(\delta v_{3}) & V_{sph.}(\delta v_{3})\\
\vdots&\vdots&\vdots\\
V_{thin}(\delta v_{m}) & V_{thick}(\delta v_{m}) & V_{sph.}(\delta v_{m})\\
\end{bmatrix}
\begin{bmatrix}
n_{thin}\\
n_{thick}\\
n_{sph.}
\end{bmatrix}
\\
\\
=
\begin{bmatrix}
N_{\ast}(\delta v_1)\\
N_{\ast}(\delta v_2)\\
N_{\ast}(\delta v_3)\\
\vdots\\
N_{\ast}(\delta v_m)
\end{bmatrix}
\end{eqnarray*}
where $\delta v_{1,2,...,m}$ are the chosen tangential velocity ranges.
Investigations indicate that a suitable set of velocity ranges is $30 < v_{1} < 50$kms$^{-1}$,
 $50 < v_{2} < 80$kms$^{-1}$, $80 < v_{3} < 120$kms$^{-1}$, $120 < v_{4} < 200$kms$^{-1}$ and
 $200 < v_{5} < \infty$kms$^{-1}$, 
thus sampling at finer resolution at highly populated velocities where the relative contributions
of each population change rapidly, and probing the region of pure spheroid stars at the extreme.
In short hand,
\begin{equation*}
V\textbf{n} = \textbf{N}
\end{equation*}
We assume no errors on the design matrix $V$, permitting a reasonable solution to be obtained by weighted least squares.
The weighted least squares solution for $\textbf{n}$, denoted $\textbf{\^n}$, is given by
\begin{equation*}
\textbf{\^n} = (V^{\dag}WV)^{-1}(V^{\dag} W N)
\end{equation*}
where $^{\dag}$ denotes the matrix transpose. We use a non-negative least squares algorithm to enforce the positivity constraint
on the parameters \citep{nnls}.
$W$ is the matrix of weights, which we set according to
\begin{equation*}
W
=
\begin{bmatrix}
\frac{1}{\sigma^2_{N_1}} & 0                        & 0                        & \cdots & 0\\
0                        & \frac{1}{\sigma^2_{N_2}} & 0                        &        &  \\
0                        & 0                        & \frac{1}{\sigma^2_{N_3}} &        &  \\
\vdots                   &                          &                          & \ddots  &  \\
0                        &                          &                          &        & \frac{1}{\sigma^2_{N_m}}
\end{bmatrix}
\end{equation*}
i.e. inverse variance weights, adopting Poisson statistics to estimate the noise on the observed number counts.
For discrete tangential velocity ranges, there is no covariance between adjacent number count bins and the weight matrix is diagonal.
The uncertainties on $\textbf{\^n}$~are obtained from the variance-covariance matrix $\mathcal{V}$ calculated
\begin{equation*}
\mathcal{V} = (V^{\dag}W V)^{-1}
\end{equation*}
and we adopt these as the formal uncertainties on the LF points. Note that the use of theroretical mean absolute
magnitudes removes the horizontal error bars.
\subsection{The luminosity functions for the disks and spheroid}
\label{decompLF}
\begin{figure*}
\begin{minipage}{160mm}
\begin{tabular}{ll}
\hspace{-1.0cm}\includegraphics[width=8.5cm]{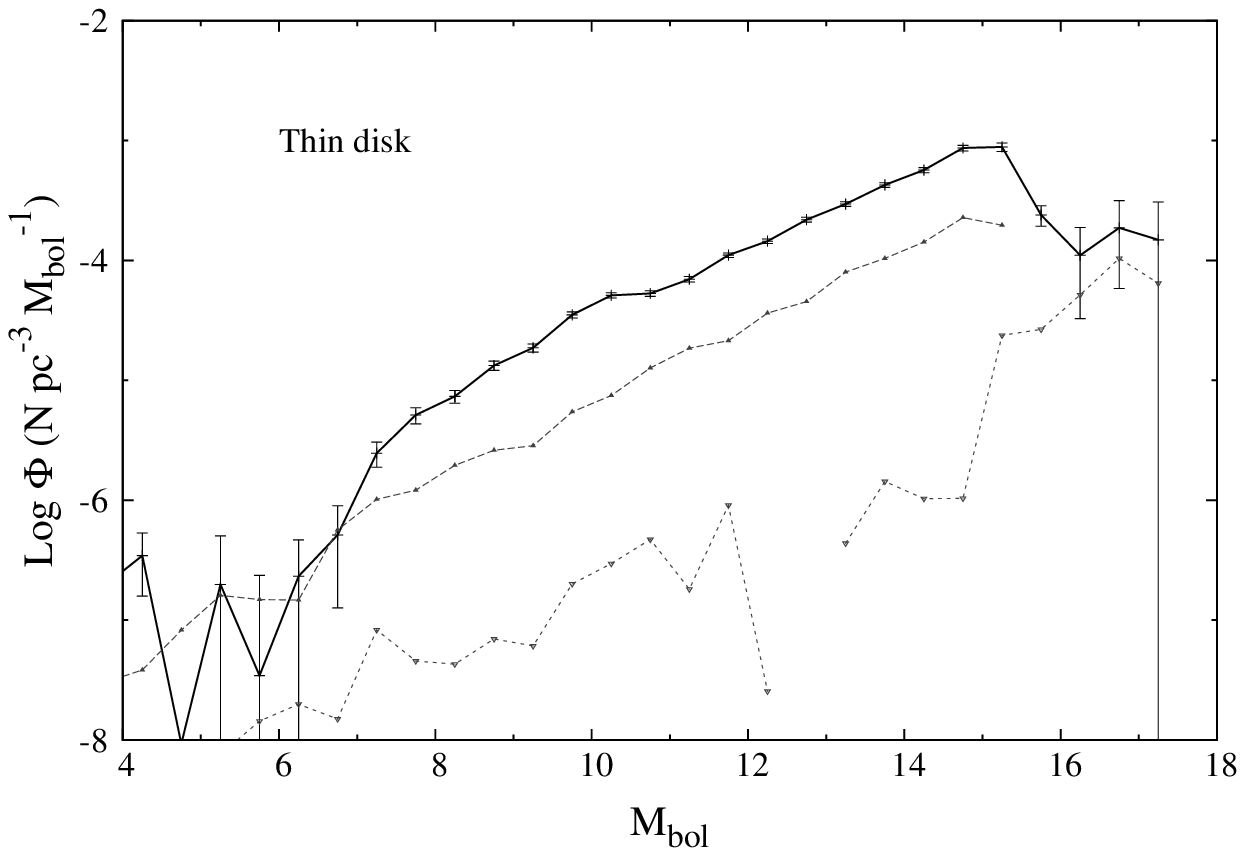}&
\hspace{-0.0cm}\includegraphics[width=8.5cm]{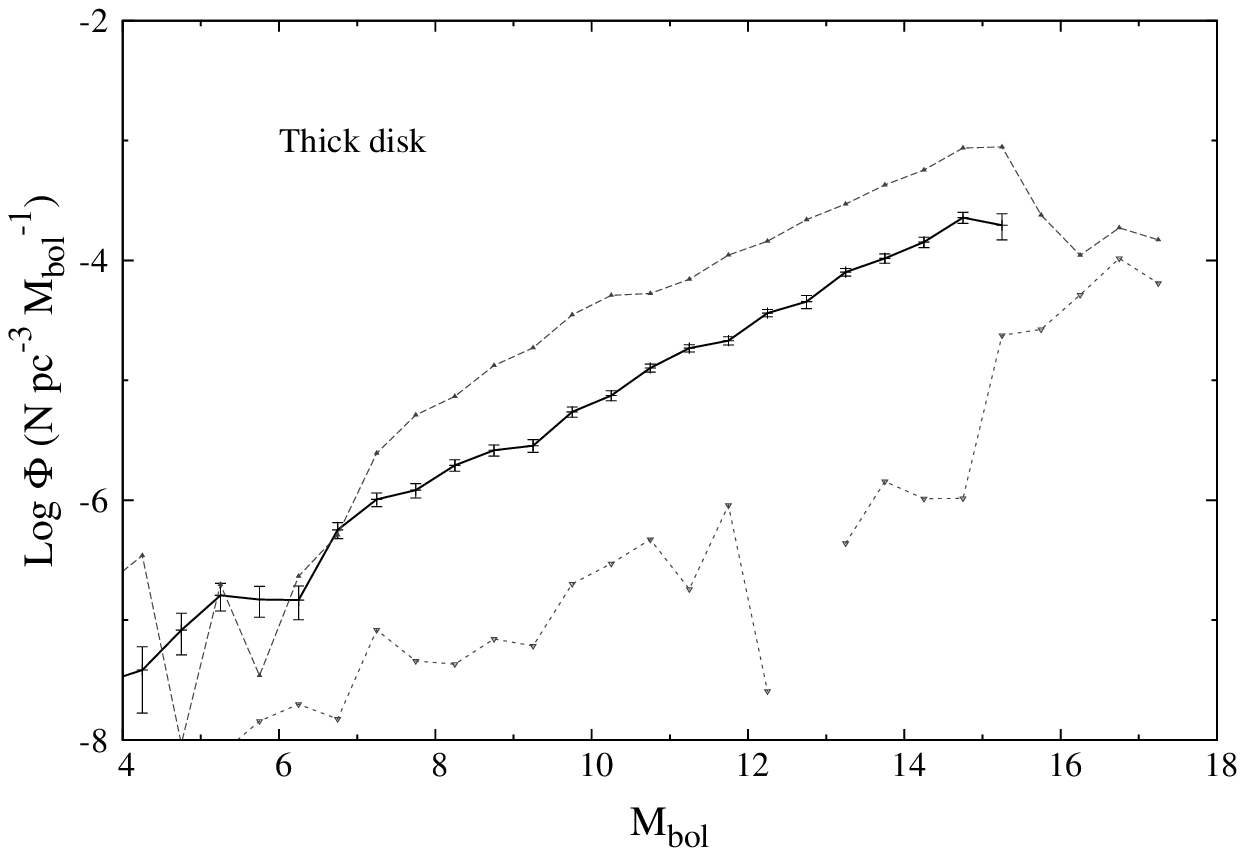}\\
\hspace{-1.0cm}\includegraphics[width=8.5cm]{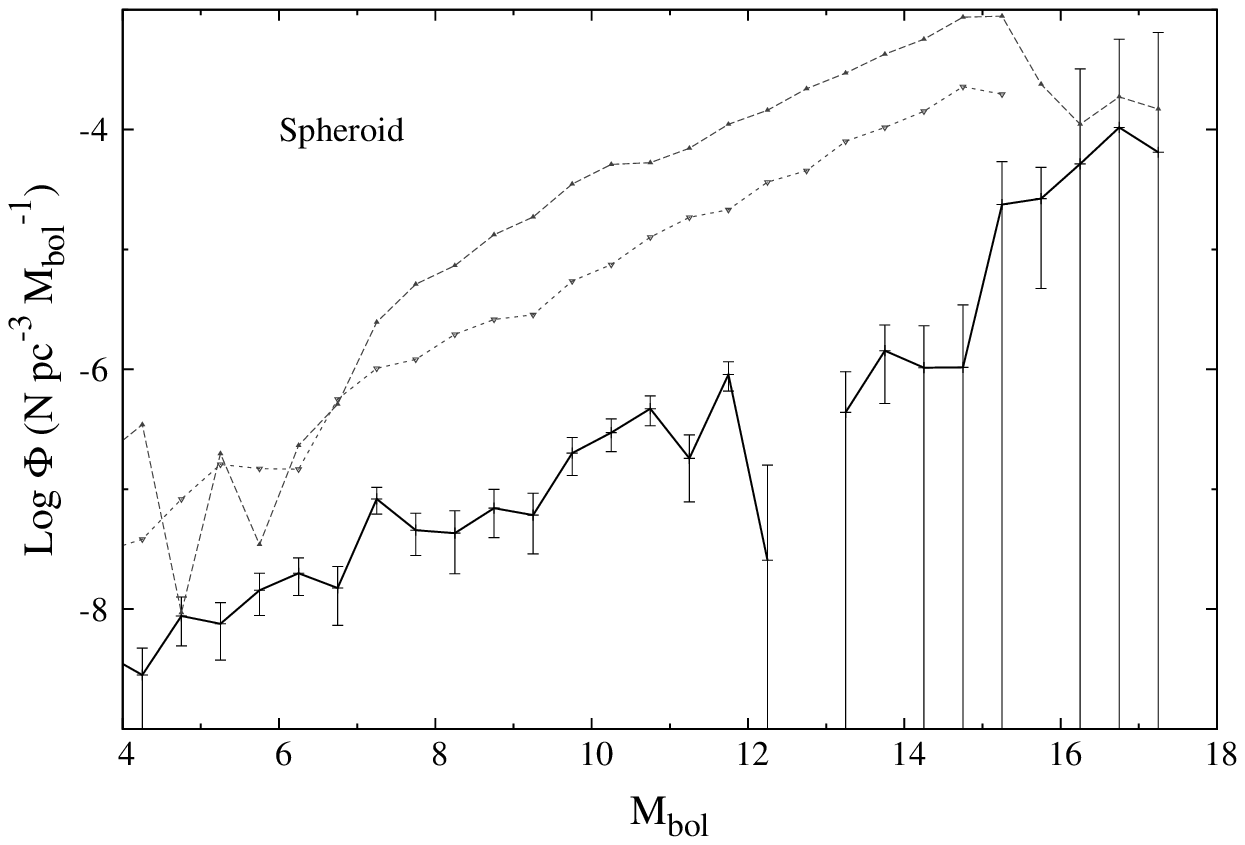}&
\hspace{-0.0cm}\includegraphics[width=8.5cm]{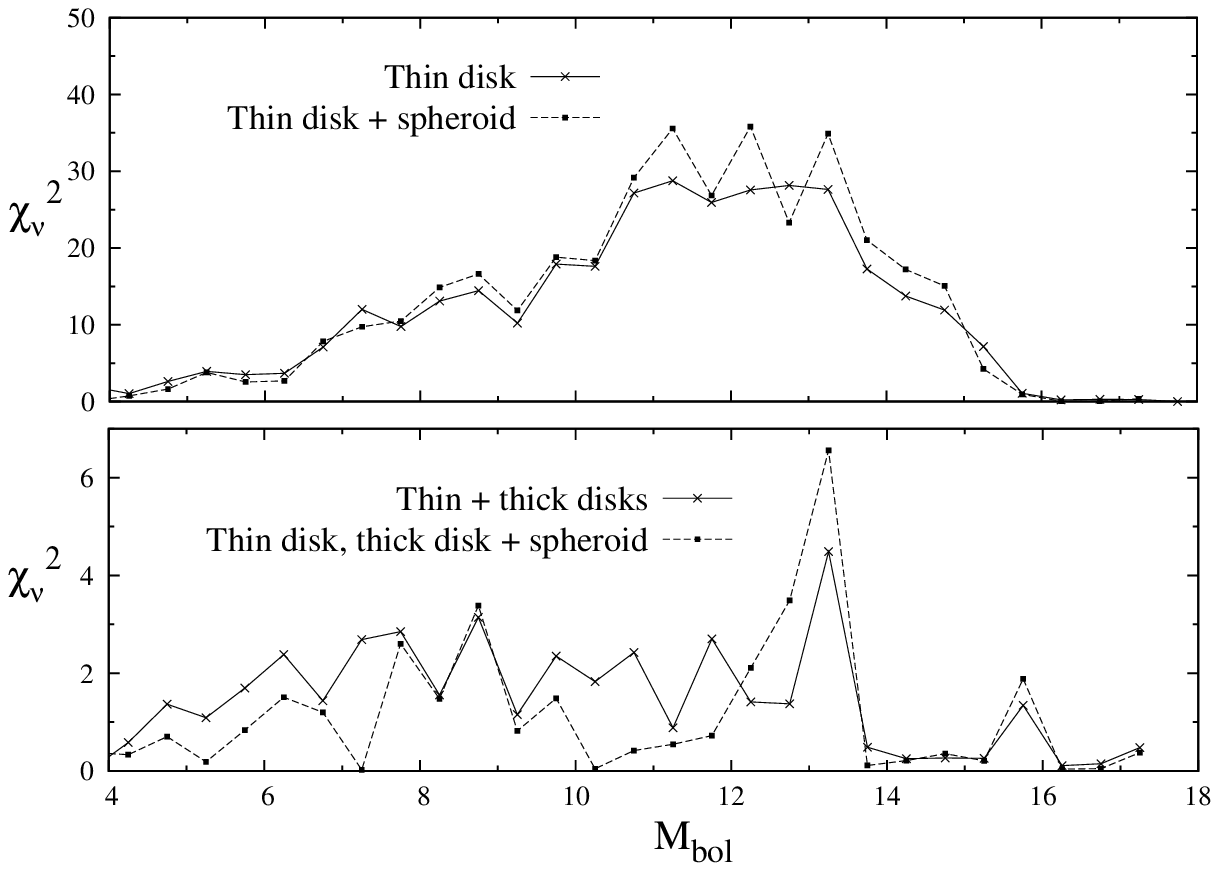}\\
\end{tabular}
\caption[Decomposed luminosity functions]{The luminosity functions for thin disk, thick disk and spheroid white dwarfs, obtained from our catalogue using the effective volume method. The lower right tile shows the $\chi^2$ statistic for each luminosity bin, using different models to predict the observed star counts.}
\label{fig:decompLFs}
\end{minipage}
\end{figure*}
The luminosity functions for the thin disk, thick disk and spheroid obtained on applying this technique to our white dwarf
catalogue are presented in Figure \ref{fig:decompLFs}. Note that we do not connect up luminosity function points across empty 
magnitude bins, where no solution was found for one or more of the populations.
The turnover in the thin disk is apparent at around $M_{bol}=15.75$,
in agreement with other works. In contrast, the spheroid luminosity function continues to rise to the faintest detected magnitudes.
No solution is found for the spheroid in a single bright magnitude bin at $M_{bol}=12.75$, which is most likely due to statistical
fluctuations in the spheroid number counts rather than a failure of the model.

The thick disk emerges nicely as an intermediate density population 
between the thin and thick disks. There is some indication of a turnover at faint magnitudes, but again the 
constraint is too poor to draw firm conclusions and beyond $M_{bol}=15.25$ this population is lost altogether.
The fact the thin and thick disks reach approximately equal density at magnitudes brighter than $M_{bol}\sim7$ is interesting, and may
hold information about the relative star formation histories of these populations. However, the $\chi_{\nu}^2$ statistic presented in 
the lower right tile indicates that a two component model excluding the thick disk would fit the data only marginally worse than
the three component model at these magnitudes, so the detection of the thick disk as a distinct entity is not conclusive. It may
be the case that the scaleheight or velocity ellipsoid of the thick disk white dwarfs is closer to that of the thin disk at these magnitudes.

We applied this technique to our white dwarf catalogue several times using different combinations of the kinematic
populations, to check that the model required all three populations to fit the data. The presence of the spheroid stars can be 
proved by conventional techniques (c.f.~Section \ref{spheroidLF}), but the thick disk can only be detected by looking for a
significant improvement in the fit when it is included in the model. The upper $\chi_{\nu}^2$ plot indicates that models with just
a thin disk, or with a thin disk and spheroid, fit the data very poorly. In the lower panel we have included the thick disk; the 
fit is clearly much better, so we regard this as the first detection of the thick disk white dwarf luminosity function. The fit
is improved marginally by the inclusion of the spheroid population.

The faintest few bins are very poorly constrained, and are significantly more uncertain than the constraint achieved using
the traditional $\frac{1}{V_{max}}$ method. This is disappointing because good constraint at these magnitudes
is necessary for accurate age measurement. Most of the age information is contained in the depth and position of the turnover at faint
magnitudes, although quantitative conclusions require analysis in conjunction with theoretical luminosity functions.
The three component model also fails at these magnitudes as no positive solution is found for the thick disk. This means the densities
recovered for the thin disk and spheroid are somewhat compromised. The reason for this may be due to the fact the faint bins are 
dominated by the ultracool white dwarfs, which have particularly uncertain photometric parallaxes and bolometric magnitudes. It may also
be the case that the both the velocity ellipsoid and scaleheight of the faintest disk white dwarfs are inflated relative to brighter 
stars, due to the faint stars being on average older. This would result in incorrect effective volumes and an inappropriate model.
\begin{figure}
\centering
\subfigure[]{\includegraphics[width=8cm]{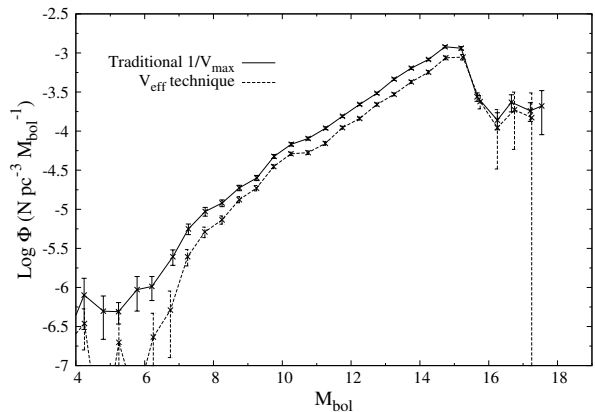}\label{fig:compDiskLSQ}}\\
\subfigure[]{\includegraphics[width=8cm]{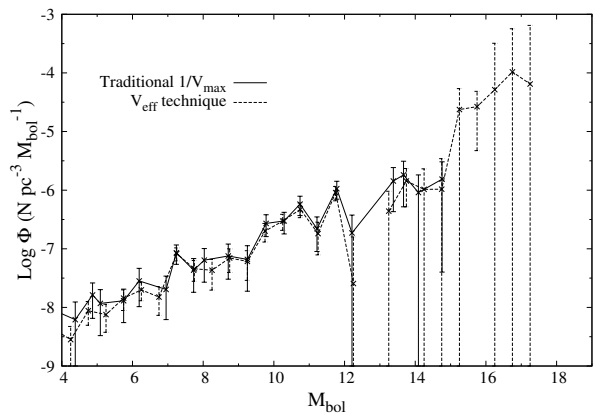}\label{fig:compSphLSQ}}
\caption[]{Comparison between LFs obtained by traditional $\frac{1}{V_{max}}$ and our effective volume technique. In the case of the
thin disk (\ref{fig:compDiskLSQ}), the lower density found by the effective volume technique is due to the fact thick disk objects have
been excluded. Both techniques agree well on the spheroid luminosity function because a clean sample can be isolated for the $\frac{1}{V_{max}}$
method. However, the effective volume technique has a smaller error because the spheroid stars at low velocity are included in the solution.}
\label{fig:comparisonLSQVmax}
\end{figure}
In Figure \ref{fig:comparisonLSQVmax}, we compare the thin disk and spheroid luminosity function obtained using the effective volume approach
to those obtained from the same star catalogue using the conventional $\frac{1}{V_{max}}$ technique.  In the case of the
thin disk (\ref{fig:compDiskLSQ}), the lower density found by the effective volume technique is due to the fact thick disk objects have
been excluded. The traditional $\frac{1}{V_{max}}$ technique used in previous studies of the disk white dwarf luminosity function does not 
distinguish thin and thick disk stars, so in this case and all others the `thin disk' luminosity function is really a sum of both the
thin and thick disks. Both techniques agree well on the spheroid luminosity function, because a clean sample can be isolated for the 
$\frac{1}{V_{max}}$ method. However, the effective volume technique has smaller error bars on all points because the spheroid stars at low 
velocity are included in the solution. It is also able to probe 2.5 magnitudes deeper, although only one of these bins has reasonable 
constraint on the density.
\subsection{Thin disk/thick disk/spheroid normalisation}
Integrating the luminosity functions in Figure \ref{fig:decompLFs} allows us to measure the total and relative densities of the
different kinematic populations in the Solar neighbourhood. For the thin disk, we find a total density of
$(2.23 \pm 0.17)\times10^{-3}$pc$^{-3}$. This is somewhat smaller than that measured in Section \ref{totalDensity}, due to the
exclusion of thick disk objects here\footnote{Note that the thick disk density is not equal to the difference, 
because in Section \ref{totalDensity} the $V_{max}$ for each thick disk object is calculated using the thin disk density profile 
and velocity ellipsoid.}. 
For the thick disk, we find a total density of $(4.55 \pm 0.41)\times10^{-4}$pc$^{-3}$, and for the spheroid we find 
$(1.4 \pm 5.6)\times10^{-4}$pc$^{-3}$. The spheroid density is considerably larger than that of Section \ref{totalDensity} due to
the fact the luminosity function extends to much fainter magnitudes where the density is greater. However the same caveat applies;
the spheroid WDLF may extend to considerably fainter magnitudes than probed here, so the density should be regarded as a lower limit. 

We can correct these values for the incompleteness in our survey if we assume the difference between the disk density
measured in Section \ref{totalDensity} and that measured by H06 is due to the incompleteness. This
boosts the densities to $3.1\times10^{-3}$pc$^{-3}$, $6.4\times10^{-4}$pc$^{-3}$ and $1.9\times10^{-4}$pc$^{-3}$
for the thin disk/thick disk/spheroid
%
%
Using these numbers, the total local density of white dwarfs is split among the kinematic populations in roughly $79\%/16\%/5\%$.
This agrees well with predictions based on Galactic models, in particular the results
of \citet{reid2005} who adopts numbers similar to these to successfully reproduce the results of a number of proper motion
surveys for white dwarfs. The similarity between Reid's Figure 9 and our luminosity functions in Figure \ref{fig:decompLFs}
is striking.
\section{Comparison to other works}
\label{eight}
\begin{figure}
\centering
\subfigure[]{\includegraphics[width=8cm]{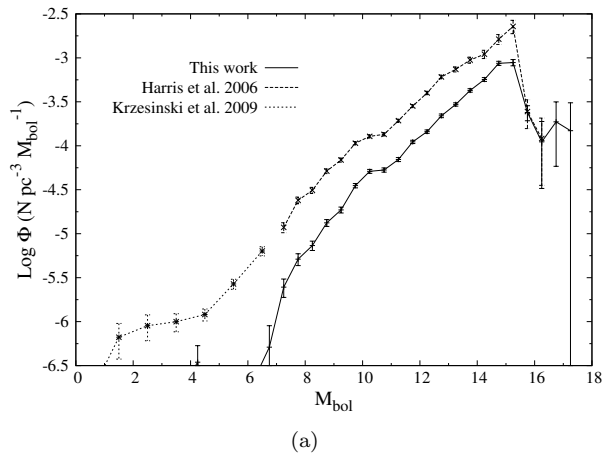}\label{fig:compDisk}}\\
\subfigure[]{\includegraphics[width=8cm]{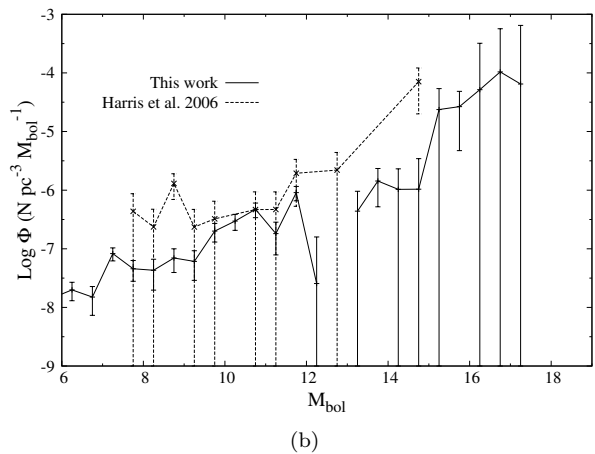}\label{fig:compSph}}
\caption[Comparison to other luminosity function studies]{Figure (a) compares the decomposed thin 
disk luminosity function derived in this work with that of H06
and \citet{krzesinski2009} who use traditional methods. The vertical offset is a combination of uniform incompleteness
and the ejection of thick disk objects from our luminosity function.
Figure (b) shows the H06 luminosity function for white dwarfs with $v_t>200$kms$^{-1}$, as well as our 
decomposed spheroid luminosity function.}
\label{fig:comparison}
\end{figure}

The most directly comparable white dwarf luminosity function to this study, in terms of number of stars and survey 
technique, is that of H06. 
These authors used proper motions derived from a combination of SDSS and USNO-B 
astrometry to obtain a sample of 6000 white dwarfs with $v_t > 30$kms$^{-1}$, with photometric parallaxes obtained
from superior 5 band SDSS photometry. Their disk luminosity function is shown 
in Figure \ref{fig:compDisk}, alongside our own reproduced from Figure \ref{fig:decompLFs}. At the bright end, the
luminosity function of \citet{krzesinski2009} is shown, also obtained from the SDSS but using selection based on
UV excess to assemble a sample of hot white dwarfs.
The most obvious difference in the thin disk luminosity functions (top panel) is the rather large vertical offset, 
which is due to a combination of the incompleteness
in our survey, and the fact that our luminosity function is exclusively for thin disk objects whereas those shown
for comparison contain thick disk objects as well. At the bright end the luminosity functions diverge, with ours
finding a considerably lower density for hot white dwarfs. Our luminosity function obtained
using the $\frac{1}{V_{max}}$ method agrees much better (e.g.~Figure \ref{fig:compDiskLSQ}), so the difference may be 
due to a genuine change in the relative contributions of the thin and thick disks at these luminosities. However, 
Section \ref{decompLF} suggests the thin and thick disk decomposition may be unreliable in this range.
There are some striking similarities between these luminosity functions too. In particular, they agree very well on 
the position of the turn over at the faint end, a feature arising from the finite age of the Galaxy.
Our $\frac{1}{V_{max}}$ luminosity function finds the same result, while achieving greater constraint.
The inflexion at $M_{bol}\sim11$ identified by H06 seems to be confirmed by this study.
This feature may hold valuable information about the recent star formation history of the disk.

The spheroid luminosity functions (lower panel) show the same vertical offset, though this is less significant
due to the larger errors. Morphological features of note include the slope of the luminosity function at brighter magnitudes
where the constraint is good, and the rather more uncertain upturn at around $M_{bol}\sim15$. Both surveys seem to agree
on these features. Our spheroid luminosity function has considerably smaller error bars at magnitudes where
the surveys overlap, and probes two and a half magnitudes fainter.
\section{Conclusions}
\label{nine}
\subsection{Proper motion survey and the luminosity function for white dwarfs}
We have conducted a wide-angle survey for white dwarfs using data from the SuperCOSMOS Sky Survey
and reduced proper motion selection. Our catalogue contains 9749 WD candidates on applying a 
$v_{\mathrm{tan}}>20$kms$^{-1}$ threshold, with photometric distances accurate to around 50\%.
Our catalogue suffers from $\sim50\%$ incompleteness due to
the ejection of blended objects and those undetected at the first epoch, however this has been shown to be
uniform and does not bias the survey. Although the normalisation of the resulting luminosity functions
is compromised their morphology is not, and any cosmochronology measurements using luminosity function models
will likewise be unaffected.

We have introduced a new technique to decompose a single mixed catalogue of stars into 
the contributions from each kinematic component, and used this to measure the thin disk, thick disk
and spheroid WDLFs separately. This is the first direct measurement of the thick disk WDLF. Our thin
disk and spheroid WDLFs probe 1.0 and 2.5 magnitudes deeper than the next deepest study (H06).
By integrating the WDLFs we have measured the relative contributions to the total WD density in the Solar neighbourhood
from each of the kinematic populations, finding results in good agreement with predictions.

The excellent agreement between our thin disk WDLF and that of H06 confirms the location of the faint turnover
at $M_{bol}=15.75$. This is surely now a secure result. An interesting feature in the rising slope of the thin disk WDLF
that is also observed by H06 would seem to be confirmed by our results. At $M_{bol}\sim11$ WD cooling times are 
of the order 500 Myr, so this feature may arise from recent star formation activity in the disk. 

%
\subsection{Decomposing the kinematic populations in the Solar neighbourhood}
Our work presents a useful way to decompose the otherwise indistinguishable WD components of the Galactic kinematic 
populations in the Solar neighbourhood, although our survey does not probe faint enough magnitudes to reach the WDLF 
terminus in each case. This technique may be applied to other types of star, although it is of particular application
in WD studies due to the fact that metallicity cannot be used to assign population membership, and the lack of 
spectral lines in cool stars means radial velocities are not available.

As our method separates the populations based on their tangential velocity distributions, the fact that there has been no rigorous study 
of the kinematics of faint white dwarfs means our velocity dispersions (adopted from main sequence stars)
must be considered a source of error. For example, assuming a thin disk velocity ellipsoid that is narrower than the true value
may result in the thick disk being over populated in our models. It may also be the case that
WDs at the faint end of the thin disk luminosity function have undergone significantly more kinematic heating than
those at brighter magnitudes, and no single velocity dispersion is truly representative. Errors on the photometric
parallaxes will also effect the partitioning of stars among the kinematic populations, as they propagate
directly into errors on the tangential velocity. The magnitude of these effects could be estimated using suitable 
Monte Carlo models, however we have not done this here.

We also point out that while the moments of the velocity distribution can be recovered from proper motions alone
\citep[e.g.~][]{dehnen1998}, reduced proper motion selected catalogues such as our own are most likely useless
for measuring velocity distributions, due to the fact that the catalogue excludes low velocity stars.
\subsection{A note on WDLF age estimates}
The technique of dating a population of stars by measuring the position of the WDLF turnover has been applied
successfully to young Galactic clusters.
However, the position of the turnover is relatively insensitive to age at the canonical disk age of 8 Gyr, 
and it is the depth and structure beyond the turnover that is the main
source of age constraint in this regime. This structure is still extremely uncertain for several reasons, including
statistical error arising from the small numbers of stars and uncertain bolometric magnitudes for the ultracool
white dwarfs that populate this range. Our work has shown that the different Galactic kinematic populations
overlap significantly beyond the thin disk LF peak, and traditional approaches to the WDLF that do not allow for the possibility of
a significant contribution from thick disk and spheroid stars cannot hope to measure an accurate thin disk age.
For these reasons it is our opinion that existing Galactic disk age measurements based on the WDLF turnover should be 
treated with caution.

Accurate aging of the kinematic populations separately would indeed be a major achievement of WD cosmochronology.
This may have to wait for results from the PanSTARRS
 or Gaia 
surveys before the observational resources are in place. These would
have to be matched by an improved understanding of the evolution and cooling times of the faintest white dwarfs, so that
some of the current systematic errors (e.g.~the spectral evolution problem) can be reduced.
\section*{Acknowledgments}
NR acknowledges the support of an STFC postgraduate studentship, which funded the majority of this work.
%
%
\bibliographystyle{mn2e}
\bibliography{references}

\begin{thebibliography}{}

\bibitem[\protect\citeauthoryear{{Bedin}, {King}, {Anderson}, {Piotto},
  {Salaris}, {Cassisi} \& {Serenelli}}{{Bedin} et~al.}{2008}]{bedin2008a}
{Bedin} L.~R.,  {King} I.~R.,  {Anderson} J.,  {Piotto} G.,  {Salaris} M.,
  {Cassisi} S.,    {Serenelli} A.,  2008, \apj, 678, 1279

\bibitem[\protect\citeauthoryear{{Bedin}, {Salaris}, {King}, {Piotto},
  {Anderson} \& {Cassisi}}{{Bedin} et~al.}{2010}]{bedin2010}
{Bedin} L.~R.,  {Salaris} M.,  {King} I.~R.,  {Piotto} G.,  {Anderson} J.,
  {Cassisi} S.,  2010, \apjl, 708, L32

\bibitem[\protect\citeauthoryear{{Bergeron} \& {Leggett}}{{Bergeron} \&
  {Leggett}}{2002}]{bergeron2002}
{Bergeron} P.,  {Leggett} S.~K.,  2002, \apj, 580, 1070

\bibitem[\protect\citeauthoryear{{Bergeron}, {Leggett} \& {Ruiz}}{{Bergeron}
  et~al.}{2001}]{BLR2001}
{Bergeron} P.,  {Leggett} S.~K.,    {Ruiz} M.~T.,  2001, \apjs, 133, 413

\bibitem[\protect\citeauthoryear{{Bergeron}, {Ruiz}, {Hamuy}, {Leggett},
  {Currie}, {Lajoie} \& {Dufour}}{{Bergeron} et~al.}{2005}]{bergeron2005}
{Bergeron} P.,  {Ruiz} M.~T.,  {Hamuy} M.,  {Leggett} S.~K.,  {Currie} M.~J.,
  {Lajoie} C.,    {Dufour} P.,  2005, \apj, 625, 838

\bibitem[\protect\citeauthoryear{{Bessell}}{{Bessell}}{1986}]{bessel1986}
{Bessell} M.~S.,  1986, \pasp, 98, 1303

\bibitem[\protect\citeauthoryear{{Chiba} \& {Beers}}{{Chiba} \&
  {Beers}}{2000}]{chiba2000}
{Chiba} M.,  {Beers} T.~C.,  2000, \aj, 119, 2843

\bibitem[\protect\citeauthoryear{{De Gennaro}, {von Hippel}, {Winget},
  {Kepler}, {Nitta}, {Koester} \& {Althaus}}{{De Gennaro}
  et~al.}{2008}]{degennaro2008}
{De Gennaro} S.,  {von Hippel} T.,  {Winget} D.~E.,  {Kepler} S.~O.,  {Nitta}
  A.,  {Koester} D.,    {Althaus} L.,  2008, \aj, 135, 1

\bibitem[\protect\citeauthoryear{{Dehnen} \& {Binney}}{{Dehnen} \&
  {Binney}}{1998}]{dehnen1998}
{Dehnen} W.,  {Binney} J.~J.,  1998, \mnras, 298, 387

\bibitem[\protect\citeauthoryear{{Eisenstein}, {Liebert}, {Harris}, {Kleinman},
  {Nitta}, {Silvestri}, {Anderson}, {Barentine}, {Brewington}, {Brinkmann},
  {Harvanek}, {Krzesi{\'n}ski}, {Neilsen} Jr., {Long}, {Schneider} \&
  {Snedden}}{{Eisenstein} et~al.}{2006}]{eisenstein2006}
{Eisenstein} D.~J.,  {Liebert} J.,  {Harris} H.~C.,  {Kleinman} S.~J.,  {Nitta}
  A.,  {Silvestri} N.,  {Anderson} S.~A.,  {Barentine} J.~C.,  {Brewington}
  H.~J.,  {Brinkmann} J.,  {Harvanek} M.,  {Krzesi{\'n}ski} J.,  {Neilsen} Jr.
  E.~H.,  {Long} D.,  {Schneider} D.~P.,    {Snedden} S.~A.,  2006, \apjs, 167,
  40

\bibitem[\protect\citeauthoryear{{Evans}}{{Evans}}{1989}]{evans1989}
{Evans} D.~W.,  1989, \aaps, 78, 249

\bibitem[\protect\citeauthoryear{{Fontaine}, {Brassard} \&
  {Bergeron}}{{Fontaine} et~al.}{2001}]{fontaine2001}
{Fontaine} G.,  {Brassard} P.,    {Bergeron} P.,  2001, \pasp, 113, 409

\bibitem[\protect\citeauthoryear{{Fuchs}, {Dettbarn}, {Rix}, {Beers},
  {Bizyaev}, {Brewington}, {Jahrei{\ss}}, {Klement}, {Malanushenko},
  {Malanushenko}, {Oravetz}, {Pan}, {Simmons} \& {Snedden}}{{Fuchs}
  et~al.}{2009}]{fuchs2009}
{Fuchs} B.,  {Dettbarn} C.,  {Rix} H.,  {Beers} T.~C.,  {Bizyaev} D.,
  {Brewington} H.,  {Jahrei{\ss}} H.,  {Klement} R.,  {Malanushenko} E.,
  {Malanushenko} V.,  {Oravetz} D.,  {Pan} K.,  {Simmons} A.,    {Snedden} S.,
  2009, \aj, 137, 4149

\bibitem[\protect\citeauthoryear{{Gates}, {Gyuk}, {Harris}, {Subbarao},
  {Anderson}, {Kleinman}, {Liebert}, {Brewington}, {Brinkmann}, {Harvanek},
  {Krzesinski}, {Lamb}, {Long}, {Neilsen} Jr., {Newman}, {Nitta} \&
  {Snedden}}{{Gates} et~al.}{2004}]{gates2004}
{Gates} E.,  {Gyuk} G.,  {Harris} H.~C.,  {Subbarao} M.,  {Anderson} S.,
  {Kleinman} S.~J.,  {Liebert} J.,  {Brewington} H.,  {Brinkmann} J.,
  {Harvanek} M.,  {Krzesinski} J.,  {Lamb} D.~Q.,  {Long} D.,  {Neilsen} Jr.
  E.~H.,  {Newman} P.~R.,  {Nitta} A.,    {Snedden} S.~A.,  2004, \apjl, 612,
  L129

\bibitem[\protect\citeauthoryear{{Gehrels}}{{Gehrels}}{1986}]{gehrels1986}
{Gehrels} N.,  1986, \apj, 303, 336

\bibitem[\protect\citeauthoryear{{Geijo}, {Torres}, {Isern} \&
  {Garc{\'{\i}}a-Berro}}{{Geijo} et~al.}{2006}]{geijo2006}
{Geijo} E.~M.,  {Torres} S.,  {Isern} J.,    {Garc{\'{\i}}a-Berro} E.,  2006,
  \mnras, 369, 1654

\bibitem[\protect\citeauthoryear{{Hambly}, {Davenhall}, {Irwin} \&
  {MacGillivray}}{{Hambly} et~al.}{2001}]{hambly2001c}
{Hambly} N.~C.,  {Davenhall} A.~C.,  {Irwin} M.~J.,    {MacGillivray} H.~T.,
  2001, \mnras

\bibitem[\protect\citeauthoryear{{Hambly}, {Henry}, {Subasavage}, {Brown} \&
  {Jao}}{{Hambly} et~al.}{2004}]{hambly2004}
{Hambly} N.~C.,  {Henry} T.~J.,  {Subasavage} J.~P.,  {Brown} M.~A.,    {Jao}
  W.-C.,  2004, \aj, 128, 437

\bibitem[\protect\citeauthoryear{{Hambly}, {Irwin} \& {MacGillivray}}{{Hambly}
  et~al.}{2001}]{hambly2001b}
{Hambly} N.~C.,  {Irwin} M.~J.,    {MacGillivray} H.~T.,  2001, \mnras

\bibitem[\protect\citeauthoryear{{Hambly}, {MacGillivray}, {Read}, {Tritton},
  {Thomson}, {Kelly}, {Morgan}, {Smith}, {Driver}, {Williamson}, {Parker},
  {Hawkins}, {Williams} \& {Lawrence}}{{Hambly} et~al.}{2001}]{hambly2001a}
{Hambly} N.~C.,  {MacGillivray} H.~T.,  {Read} M.~A.,  {Tritton} S.~B.,
  {Thomson} E.~B.,  {Kelly} B.~D.,  {Morgan} D.~H.,  {Smith} R.~E.,  {Driver}
  S.~P.,  {Williamson} J.,  {Parker} Q.~A.,  {Hawkins} M.~R.~S.,  {Williams}
  P.~M.,    {Lawrence} A.,  2001, \mnras

\bibitem[\protect\citeauthoryear{{Hansen}}{{Hansen}}{1999}]{hansen1999}
{Hansen} B.~M.~S.,  1999, \apj, 520, 680

\bibitem[\protect\citeauthoryear{{Hansen}, {Brewer}, {Fahlman}, {Gibson},
  {Ibata}, {Limongi}, {Rich}, {Richer}, {Shara} \& {Stetson}}{{Hansen}
  et~al.}{2002}]{hansen2002}
{Hansen} B.~M.~S.,  {Brewer} J.,  {Fahlman} G.~G.,  {Gibson} B.~K.,  {Ibata}
  R.,  {Limongi} M.,  {Rich} R.~M.,  {Richer} H.~B.,  {Shara} M.~M.,
  {Stetson} P.~B.,  2002, \apjl, 574, L155

\bibitem[\protect\citeauthoryear{{Hansen} \& {Liebert}}{{Hansen} \&
  {Liebert}}{2003}]{hansen2003}
{Hansen} B.~M.~S.,  {Liebert} J.,  2003, \araa, 41, 465

\bibitem[\protect\citeauthoryear{{Harris}, {Dahn}, {Vrba}, {Henden}, {Liebert},
  {Schmidt} \& {Reid}}{{Harris} et~al.}{1999}]{harris1999}
{Harris} H.~C.,  {Dahn} C.~C.,  {Vrba} F.~J.,  {Henden} A.~A.,  {Liebert} J.,
  {Schmidt} G.~D.,    {Reid} I.~N.,  1999, \apj, 524, 1000

\bibitem[\protect\citeauthoryear{{Harris}, {Gates}, {Gyuk}, {Subbarao},
  {Anderson}, {Hall}, {Munn}, {Liebert}, {Knapp}, {Bizyaev}, {Malanushenko},
  {Malanushenko}, {Pan}, {Schneider} \& {Smith}}{{Harris}
  et~al.}{2008}]{harris2008}
{Harris} H.~C.,  {Gates} E.,  {Gyuk} G.,  {Subbarao} M.,  {Anderson} S.~F.,
  {Hall} P.~B.,  {Munn} J.~A.,  {Liebert} J.,  {Knapp} G.~R.,  {Bizyaev} D.,
  {Malanushenko} E.,  {Malanushenko} V.,  {Pan} K.,  {Schneider} D.~P.,
  {Smith} J.~A.,  2008, \apj, 679, 697

\bibitem[\protect\citeauthoryear{{Harris}, {Munn}, {Kilic}, {Liebert},
  {Williams}, {von Hippel}, {Levine}, {Monet}, {Eisenstein}, {Kleinman} \& {et
  al.}}{{Harris} et~al.}{2006}]{harris2006}
{Harris} H.~C.,  {Munn} J.~A.,  {Kilic} M.,  {Liebert} J.,  {Williams} K.~A.,
  {von Hippel} T.,  {Levine} S.~E.,  {Monet} D.~G.,  {Eisenstein} D.~J.,
  {Kleinman} S.~J.,    {et al.} 2006, \aj, 131, 571

\bibitem[\protect\citeauthoryear{{Holberg}}{{Holberg}}{2009}]{holberg2009}
{Holberg} J.~B.,  2009, Journal of Physics Conference Series, 172, 012022

\bibitem[\protect\citeauthoryear{{Jones}, {Fong}, {Shanks}, {Ellis} \&
  {Peterson}}{{Jones} et~al.}{1991}]{jones1991}
{Jones} L.~R.,  {Fong} R.,  {Shanks} T.,  {Ellis} R.~S.,    {Peterson} B.~A.,
  1991, \mnras, 249, 481

\bibitem[\protect\citeauthoryear{{Kilic}, {Leggett}, {Tremblay}, {von Hippel},
  {Bergeron}, {Harris}, {Munn}, {Williams}, {Gates} \& {Farihi}}{{Kilic}
  et~al.}{2010}]{kilic2010UCWDs}
{Kilic} M.,  {Leggett} S.~K.,  {Tremblay} P.,  {von Hippel} T.,  {Bergeron} P.,
   {Harris} H.~C.,  {Munn} J.~A.,  {Williams} K.~A.,  {Gates} E.,    {Farihi}
  J.,  2010, \apjs, 190, 77

\bibitem[\protect\citeauthoryear{{Kilic}, {Munn}, {Harris}, {Liebert}, {von
  Hippel}, {Williams}, {Metcalfe}, {Winget} \& {Levine}}{{Kilic}
  et~al.}{2006}]{kilic2006}
{Kilic} M.,  {Munn} J.~A.,  {Harris} H.~C.,  {Liebert} J.,  {von Hippel} T.,
  {Williams} K.~A.,  {Metcalfe} T.~S.,  {Winget} D.~E.,    {Levine} S.~E.,
  2006, \aj, 131, 582

\bibitem[\protect\citeauthoryear{{Kilic}, {Munn}, {Williams}, {Kowalski}, {von
  Hippel}, {Harris}, {Jeffery}, {DeGennaro}, {Brown} \& {McLeod}}{{Kilic}
  et~al.}{2010}]{kilic2010halo}
{Kilic} M.,  {Munn} J.~A.,  {Williams} K.~A.,  {Kowalski} P.~M.,  {von Hippel}
  T.,  {Harris} H.~C.,  {Jeffery} E.~J.,  {DeGennaro} S.,  {Brown} W.~R.,
  {McLeod} B.,  2010, \apjl, 715, L21

\bibitem[\protect\citeauthoryear{{Knox}, {Hawkins} \& {Hambly}}{{Knox}
  et~al.}{1999}]{knox1999}
{Knox} R.~A.,  {Hawkins} M.~R.~S.,    {Hambly} N.~C.,  1999, \mnras, 306, 736

\bibitem[\protect\citeauthoryear{{Krzesinski}, {Kleinman}, {Nitta},
  {H{\"u}gelmeyer}, {Dreizler}, {Liebert} \& {Harris}}{{Krzesinski}
  et~al.}{2009}]{krzesinski2009}
{Krzesinski} J.,  {Kleinman} S.~J.,  {Nitta} A.,  {H{\"u}gelmeyer} S.,
  {Dreizler} S.,  {Liebert} J.,    {Harris} H.,  2009, \aap, 508, 339

\bibitem[\protect\citeauthoryear{{Lawson} \& {Hanson}}{{Lawson} \&
  {Hanson}}{1974}]{nnls}
{Lawson} C.~L.,  {Hanson} R.~J.,  1974, {Solving Least Squares Problems}.
Prentice-Hall

\bibitem[\protect\citeauthoryear{{Leggett}, {Ruiz} \& {Bergeron}}{{Leggett}
  et~al.}{1998}]{legget1998}
{Leggett} S.~K.,  {Ruiz} M.~T.,    {Bergeron} P.,  1998, \apj, 497, 294

\bibitem[\protect\citeauthoryear{{Liebert}, {Bergeron} \& {Holberg}}{{Liebert}
  et~al.}{2005}]{liebert2005}
{Liebert} J.,  {Bergeron} P.,    {Holberg} J.~B.,  2005, \apjs, 156, 47

\bibitem[\protect\citeauthoryear{{Liebert}, {Dahn} \& {Monet}}{{Liebert}
  et~al.}{1988}]{liebert1988}
{Liebert} J.,  {Dahn} C.~C.,    {Monet} D.~G.,  1988, \apj, 332, 891

\bibitem[\protect\citeauthoryear{{McMillan} \& {Binney}}{{McMillan} \&
  {Binney}}{2009}]{mcmillan2009}
{McMillan} P.~J.,  {Binney} J.~J.,  2009, \mnras, 400, L103

\bibitem[\protect\citeauthoryear{{Mendez} \& {Guzman}}{{Mendez} \&
  {Guzman}}{1998}]{mendez1998}
{Mendez} R.~A.,  {Guzman} R.,  1998, \aap, 333, 106

\bibitem[\protect\citeauthoryear{{Metcalfe}, {Shanks}, {Fong} \&
  {Jones}}{{Metcalfe} et~al.}{1991}]{metcalfe1991}
{Metcalfe} N.,  {Shanks} T.,  {Fong} R.,    {Jones} L.~R.,  1991, \mnras, 249,
  498

\bibitem[\protect\citeauthoryear{{Murray}}{{Murray}}{1983}]{murray1983}
{Murray} C.~A.,  1983, {Vectorial Astrometry}.
Taylor \& Francis

\bibitem[\protect\citeauthoryear{{Oswalt}, {Smith}, {Wood} \&
  {Hintzen}}{{Oswalt} et~al.}{1996}]{oswalt1996}
{Oswalt} T.~D.,  {Smith} J.~A.,  {Wood} M.~A.,    {Hintzen} P.,  1996, \nat,
  382, 692

\bibitem[\protect\citeauthoryear{{Reed}}{{Reed}}{2006}]{reed2006}
{Reed} B.~C.,  2006, \jrasc, 100, 146

\bibitem[\protect\citeauthoryear{{Reid}}{{Reid}}{2005}]{reid2005}
{Reid} I.~N.,  2005, \araa, 43, 247

\bibitem[\protect\citeauthoryear{{Robin}, {Reyl{\'e}}, {Derri{\`e}re} \&
  {Picaud}}{{Robin} et~al.}{2003}]{robin2003}
{Robin} A.~C.,  {Reyl{\'e}} C.,  {Derri{\`e}re} S.,    {Picaud} S.,  2003,
  \aap, 409, 523

\bibitem[\protect\citeauthoryear{{Rowell}, {Kilic} \& {Hambly}}{{Rowell}
  et~al.}{2008}]{rowell2008}
{Rowell} N.~R.,  {Kilic} M.,    {Hambly} N.~C.,  2008, \mnras, 385, L23

\bibitem[\protect\citeauthoryear{{Ruiz} \& {Bergeron}}{{Ruiz} \&
  {Bergeron}}{2001}]{ruiz2001}
{Ruiz} M.~T.,  {Bergeron} P.,  2001, \apj, 558, 761

\bibitem[\protect\citeauthoryear{{Salim}, {Rich}, {Hansen}, {Koopmans},
  {Oppenheimer} \& {Blandford}}{{Salim} et~al.}{2004}]{salim2004}
{Salim} S.,  {Rich} R.~M.,  {Hansen} B.~M.,  {Koopmans} L.~V.~E.,
  {Oppenheimer} B.~R.,    {Blandford} R.~D.,  2004, \apj, 601, 1075

\bibitem[\protect\citeauthoryear{{Savitzky} \& {Golay}}{{Savitzky} \&
  {Golay}}{1964}]{savitzky1964}
{Savitzky} A.,  {Golay} M.~J.~E.,  1964, Analytical Chemistry, 36, 1627

\bibitem[\protect\citeauthoryear{{Schmidt}}{{Schmidt}}{1968}]{schmidt1968}
{Schmidt} M.,  1968, \apj, 151, 393

\bibitem[\protect\citeauthoryear{{Stobie}, {Ishida} \& {Peacock}}{{Stobie}
  et~al.}{1989}]{stobie1989}
{Stobie} R.~S.,  {Ishida} K.,    {Peacock} J.~A.,  1989, \mnras, 238, 709

\bibitem[\protect\citeauthoryear{{Tinney}, {Reid} \& {Mould}}{{Tinney}
  et~al.}{1993}]{tinney1993}
{Tinney} C.~G.,  {Reid} I.~N.,    {Mould} J.~R.,  1993, \apj, 414, 254

\bibitem[\protect\citeauthoryear{{Tremblay} \& {Bergeron}}{{Tremblay} \&
  {Bergeron}}{2008}]{tremblay2008}
{Tremblay} P.,  {Bergeron} P.,  2008, \apj, 672, 1144

\bibitem[\protect\citeauthoryear{{Vidrih}, {Bramich}, {Hewett}, {Evans},
  {Gilmore}, {Hodgkin}, {Smith}, {Wyrzykowski}, {Belokurov}, {Fellhauer} \& {et
  al.}}{{Vidrih} et~al.}{2007}]{vidrih2007}
{Vidrih} S.,  {Bramich} D.~M.,  {Hewett} P.~C.,  {Evans} N.~W.,  {Gilmore} G.,
  {Hodgkin} S.,  {Smith} M.,  {Wyrzykowski} L.,  {Belokurov} V.,  {Fellhauer}
  M.,    {et al.} 2007, \mnras, 382, 515

\bibitem[\protect\citeauthoryear{{Winget}, {Hansen}, {Liebert}, {van Horn},
  {Fontaine}, {Nather}, {Kepler} \& {Lamb}}{{Winget} et~al.}{1987}]{winget1987}
{Winget} D.~E.,  {Hansen} C.~J.,  {Liebert} J.,  {van Horn} H.~M.,  {Fontaine}
  G.,  {Nather} R.~E.,  {Kepler} S.~O.,    {Lamb} D.~Q.,  1987, \apjl, 315, L77

\end{thebibliography}

\label{lastpage}

\end{document}